\documentclass[onecolumn,superscriptaddress,secnumarabic,amssymb, nobibnotes, aps, prb]{revtex4-2}

\usepackage{amsmath}    
\usepackage{graphicx}    
\usepackage{verbatim}    
\usepackage{color}           
\usepackage{subfigure}   
\usepackage{hyperref}    
\usepackage{braket}
\usepackage{physics}
\usepackage{xfrac}
\usepackage{xspace}
\newcommand{\bbm}{\mathbf{m}}

\newcommand{\bL}{\mathbf{L}}
\newcommand{\bB}{\mathbf{B}}

\def\bL{\mathbf{L}}

\begin{document}
\title{Magnetic Dichroism in Rutile NiF$_2$: Separating Altermagnetic and Ferromagnetic Effects}

\author{A.~Hariki}
\altaffiliation{A.H. and K.S. contributed equally to this work.}
\affiliation{Department of Physics and Electronics, Graduate School of Engineering,
Osaka Metropolitan University, 1-1 Gakuen-cho, Nakaku, Sakai, Osaka 599-8531, Japan}
\author{K.~Sakurai}
\altaffiliation{A.H. and K.S. contributed equally to this work.}
\affiliation{Department of Physics and Electronics, Graduate School of Engineering,
Osaka Metropolitan University, 1-1 Gakuen-cho, Nakaku, Sakai, Osaka 599-8531, Japan}
\author{T.~Okauchi}
\affiliation{Department of Physics and Electronics, Graduate School of Engineering,
Osaka Metropolitan University, 1-1 Gakuen-cho, Nakaku, Sakai, Osaka 599-8531, Japan}
\author{J.~Kune\v{s}}
\affiliation{Department of Condensed Matter Physics, Faculty of
  Science, Masaryk University, Kotl\'a\v{r}sk\'a 2, 611 37 Brno,
  Czechia}


\begin{figure}
\caption{{\bf Orientation of magnetic moments in the (001) plane of NiF$_2$.}\\
(a) Top view of the rutile structure with the N\'eel vector $\bL$ in the [010] direction.
The red arrows mark the local moments $\bbm_{1,2}$ in the Ni sites, $\bbm$ is the net magnetization.
(b) The definitions of angles $\varphi_B$, $\varphi_L$ and $\varphi_m$ measured from $[100]$ direction
where $\bB$ in the  magnetic field applied in the $(001)$ plane. Note that the competition between 
the magneto-crystalline anisotropy and the external field leads to misalignment of net magnetization $\bbm$ and
the external magnetic field $\bB$.}
\label{fig:struct}
\end{figure}

\begin{figure}
\caption{{\bf X-ray absorption at Ni L$_{2,3}$ edges.}\\
The XAS calculated for the two circular polarizations (red and blue) at the Ni $L_{2,3}$ edge together with the XMCD intensities (shaded). The calculated spectral intensities are broadened by a Lorentzian of 0.30~eV (HWHM). The experimental Ni $L_{2,3}$-edge XAS spectrum taken from Ref.~24 is shown for comparison. The experimental baseline was offset for the sake of clarity.}
\label{fig:xmcd_exp}
\end{figure}
\begin{figure*}
\begin{center}
\end{center}
\caption{{\bf X-ray circular dichroism for magnetic field along (100) direction.}\\
(a) Ni $L_{2,3}$-edge XMCD intensities in NiF$_2$ calculated independently 
for each magnetic field $B$ and $\varphi_B = 0^\circ$ with no approximations to the method ('Full').
(b) XMCD intensities computed as a linear combination ('Approx.') of (c) $\Delta_{\rm ALT}(\omega)$ (blue, left axis) and $\Delta_{\rm FM}(\omega)$ (red, right axis) following Eq.~\ref{eq:h}. 
$\Delta_{\rm ALT}(\omega)$ in the non-relativistic limit, calculated without the Ni 3$d$ valence SOC, is also shown (thin black, left axis).
The inset in panel (b) shows the difference in the XMCD intensities at the Ni $L_3$-edge between the full calculations in (a) and the approximations in (b).}
\label{fig:xmcd1}
\end{figure*}
\begin{figure*}
\caption{{\bf Magnetic moments for the magnetic field in general orientation within the (001) plane.}\\
Calculated relation between the angles of the external field $\varphi_B$ and (a) the N\'eel vector $\varphi_L$, (b) the FM moment $\varphi_m$, and (c) the amplitude of the FM moment $|{\bf m(B)}|$ for selected amplitudes of the external field $\bf B$. The results over a wider range for the angles and amplitudes are provided in Supplementary Figure 4~[34].}
\label{fig:ang_dep}
\end{figure*}
\begin{figure*}
\caption{{\bf XMCD spectra for the magnetic field in $\varphi_B = 20^\circ$ orientation.}\\
(a) Ni $L_{2,3}$-edge XMCD intensities in NiF$_2$ calculated for various amplitudes of the magnetic field $B$ with $\varphi_B = 20^\circ$ for the two geometries of the light propagation vector $\hat{\bf k}=[100]$ (top) and $\hat{\bf k}=[010]$ (bottom). The altermagnetic contributions (b) and  the ferromagnetic contributions (c) obtained from Eq.~1
using the magnetic order parameters shown in Fig.~\ref{fig:ang_dep}. (d) The difference between 
corresponding spectra in panel (a) and the sum of panels (b)+(c).
The results for different angles $\varphi_B$ can be found in Supplementary Figure 5~[34].}
\label{fig:xmcd2}
\end{figure*}
\begin{figure}
\caption{{\bf XMCD spectra for the magnetic field parallel to the beam direction.}\\
(a) Ni $L_{2,3}$-edge XMCD intensities in NiF$_2$ calculated with the magnetic field of 10~T and varying angles $\varphi_B$. The light propagation vector $\hat{\bf k} = \hat{\bf B}$ is used in the calculation.
(b) The difference of the XMCD intensities between the full calculations in (a) and the approximations of Eq.~2 with $\alpha=0.34$ and $m_0=0.17~\mu_B$.}
\label{fig:sample_rot}
\end{figure}


\begin{thebibliography}{42}%
\makeatletter
\providecommand \@ifxundefined [1]{%
 \@ifx{#1\undefined}
}%
\providecommand \@ifnum [1]{%
 \ifnum #1\expandafter \@firstoftwo
 \else \expandafter \@secondoftwo
 \fi
}%
\providecommand \@ifx [1]{%
 \ifx #1\expandafter \@firstoftwo
 \else \expandafter \@secondoftwo
 \fi
}%
\providecommand \natexlab [1]{#1}%
\providecommand \enquote  [1]{``#1''}%
\providecommand \bibnamefont  [1]{#1}%
\providecommand \bibfnamefont [1]{#1}%
\providecommand \citenamefont [1]{#1}%
\providecommand \href@noop [0]{\@secondoftwo}%
\providecommand \href [0]{\begingroup \@sanitize@url \@href}%
\providecommand \@href[1]{\@@startlink{#1}\@@href}%
\providecommand \@@href[1]{\endgroup#1\@@endlink}%
\providecommand \@sanitize@url [0]{\catcode `\\12\catcode `\$12\catcode `\&12\catcode `\#12\catcode `\^12\catcode `\_12\catcode `\%12\relax}%
\providecommand \@@startlink[1]{}%
\providecommand \@@endlink[0]{}%
\providecommand \url  [0]{\begingroup\@sanitize@url \@url }%
\providecommand \@url [1]{\endgroup\@href {#1}{\urlprefix }}%
\providecommand \urlprefix  [0]{URL }%
\providecommand \Eprint [0]{\href }%
\providecommand \doibase [0]{https://doi.org/}%
\providecommand \selectlanguage [0]{\@gobble}%
\providecommand \bibinfo  [0]{\@secondoftwo}%
\providecommand \bibfield  [0]{\@secondoftwo}%
\providecommand \translation [1]{[#1]}%
\providecommand \BibitemOpen [0]{}%
\providecommand \bibitemStop [0]{}%
\providecommand \bibitemNoStop [0]{.\EOS\space}%
\providecommand \EOS [0]{\spacefactor3000\relax}%
\providecommand \BibitemShut  [1]{\csname bibitem#1\endcsname}%
\let\auto@bib@innerbib\@empty
\bibitem [{\citenamefont {\ifmmode~\check{S}\else \v{S}\fi{}mejkal}\ \emph {et~al.}(2022{\natexlab{a}})\citenamefont {\ifmmode~\check{S}\else \v{S}\fi{}mejkal}, \citenamefont {Sinova},\ and\ \citenamefont {Jungwirth}}]{Smejkal22a}%
  \BibitemOpen
  \bibfield  {author} {\bibinfo {author} {\bibfnamefont {L.}~\bibnamefont {\ifmmode~\check{S}\else \v{S}\fi{}mejkal}}, \bibinfo {author} {\bibfnamefont {J.}~\bibnamefont {Sinova}},\ and\ \bibinfo {author} {\bibfnamefont {T.}~\bibnamefont {Jungwirth}},\ }\bibfield  {title} {\bibinfo {title} {Emerging {R}esearch {L}andscape of {A}ltermagnetism},\ }\href {https://doi.org/10.1103/PhysRevX.12.040501} {\bibfield  {journal} {\bibinfo  {journal} {Phys. Rev. X}\ }\textbf {\bibinfo {volume} {12}},\ \bibinfo {pages} {040501} (\bibinfo {year} {2022}{\natexlab{a}})}\BibitemShut {NoStop}%
\bibitem [{\citenamefont {\ifmmode~\check{S}\else \v{S}\fi{}mejkal}\ \emph {et~al.}(2022{\natexlab{b}})\citenamefont {\ifmmode~\check{S}\else \v{S}\fi{}mejkal}, \citenamefont {Sinova},\ and\ \citenamefont {Jungwirth}}]{Smejkal22}%
  \BibitemOpen
  \bibfield  {author} {\bibinfo {author} {\bibfnamefont {L.}~\bibnamefont {\ifmmode~\check{S}\else \v{S}\fi{}mejkal}}, \bibinfo {author} {\bibfnamefont {J.}~\bibnamefont {Sinova}},\ and\ \bibinfo {author} {\bibfnamefont {T.}~\bibnamefont {Jungwirth}},\ }\bibfield  {title} {\bibinfo {title} {Beyond {C}onventional {F}erromagnetism and {A}ntiferromagnetism: {A} {P}hase with {N}onrelativistic {S}pin and {C}rystal {R}otation {S}ymmetry},\ }\href {https://doi.org/10.1103/PhysRevX.12.031042} {\bibfield  {journal} {\bibinfo  {journal} {Phys. Rev. X}\ }\textbf {\bibinfo {volume} {12}},\ \bibinfo {pages} {031042} (\bibinfo {year} {2022}{\natexlab{b}})}\BibitemShut {NoStop}%
\bibitem [{\citenamefont {Ahn}\ \emph {et~al.}(2019)\citenamefont {Ahn}, \citenamefont {Hariki}, \citenamefont {Lee},\ and\ \citenamefont {Kune\ifmmode~\check{s}\else \v{s}\fi{}}}]{Ahn19}%
  \BibitemOpen
  \bibfield  {author} {\bibinfo {author} {\bibfnamefont {K.-H.}\ \bibnamefont {Ahn}}, \bibinfo {author} {\bibfnamefont {A.}~\bibnamefont {Hariki}}, \bibinfo {author} {\bibfnamefont {K.-W.}\ \bibnamefont {Lee}},\ and\ \bibinfo {author} {\bibfnamefont {J.}~\bibnamefont {Kune\ifmmode~\check{s}\else \v{s}\fi{}}},\ }\bibfield  {title} {\bibinfo {title} {Antiferromagnetism in {RuO$_2$} as $d$-wave {Pomeranchuk} instability},\ }\href {https://doi.org/10.1103/PhysRevB.99.184432} {\bibfield  {journal} {\bibinfo  {journal} {Phys. Rev. B}\ }\textbf {\bibinfo {volume} {99}},\ \bibinfo {pages} {184432} (\bibinfo {year} {2019})}\BibitemShut {NoStop}%
\bibitem [{\citenamefont {Naka}\ \emph {et~al.}(2019)\citenamefont {Naka}, \citenamefont {Hayami}, \citenamefont {Kusunose}, \citenamefont {Yanagi}, \citenamefont {Motome},\ and\ \citenamefont {Seo}}]{Naka2019}%
  \BibitemOpen
  \bibfield  {author} {\bibinfo {author} {\bibfnamefont {M.}~\bibnamefont {Naka}}, \bibinfo {author} {\bibfnamefont {S.}~\bibnamefont {Hayami}}, \bibinfo {author} {\bibfnamefont {H.}~\bibnamefont {Kusunose}}, \bibinfo {author} {\bibfnamefont {Y.}~\bibnamefont {Yanagi}}, \bibinfo {author} {\bibfnamefont {Y.}~\bibnamefont {Motome}},\ and\ \bibinfo {author} {\bibfnamefont {H.}~\bibnamefont {Seo}},\ }\bibfield  {title} {\bibinfo {title} {Spin current generation in organic antiferromagnets},\ }\href {https://doi.org/10.1038/s41467-019-12229-y} {\bibfield  {journal} {\bibinfo  {journal} {Nature Communications}\ }\textbf {\bibinfo {volume} {10}},\ \bibinfo {pages} {4305} (\bibinfo {year} {2019})}\BibitemShut {NoStop}%
\bibitem [{\citenamefont {Hayami}\ \emph {et~al.}(2019)\citenamefont {Hayami}, \citenamefont {Yanagi},\ and\ \citenamefont {Kusunose}}]{Hayami19}%
  \BibitemOpen
  \bibfield  {author} {\bibinfo {author} {\bibfnamefont {S.}~\bibnamefont {Hayami}}, \bibinfo {author} {\bibfnamefont {Y.}~\bibnamefont {Yanagi}},\ and\ \bibinfo {author} {\bibfnamefont {H.}~\bibnamefont {Kusunose}},\ }\bibfield  {title} {\bibinfo {title} {Momentum-{D}ependent {S}pin {S}plitting by {C}ollinear {A}ntiferromagnetic {O}rdering},\ }\href {https://doi.org/10.7566/JPSJ.88.123702} {\bibfield  {journal} {\bibinfo  {journal} {J. Phys. Soc. Jpn.}\ }\textbf {\bibinfo {volume} {88}},\ \bibinfo {pages} {123702} (\bibinfo {year} {2019})}\BibitemShut {NoStop}%
\bibitem [{\citenamefont {\v{S}mejkal}\ \emph {et~al.}(2020)\citenamefont {\v{S}mejkal}, \citenamefont {Gonz\'alez-Hern\'andez}, \citenamefont {Jungwirth},\ and\ \citenamefont {Sinova}}]{Smejkal20}%
  \BibitemOpen
  \bibfield  {author} {\bibinfo {author} {\bibfnamefont {L.}~\bibnamefont {\v{S}mejkal}}, \bibinfo {author} {\bibfnamefont {R.}~\bibnamefont {Gonz\'alez-Hern\'andez}}, \bibinfo {author} {\bibfnamefont {T.}~\bibnamefont {Jungwirth}},\ and\ \bibinfo {author} {\bibfnamefont {J.}~\bibnamefont {Sinova}},\ }\bibfield  {title} {\bibinfo {title} {Crystal time-reversal symmetry breaking and spontaneous {Hall} effect in collinear antiferromagnets},\ }\href {https://doi.org/10.1126/sciadv.aaz8809} {\bibfield  {journal} {\bibinfo  {journal} {Sci. Adv.}\ }\textbf {\bibinfo {volume} {6}},\ \bibinfo {pages} {eaaz8809} (\bibinfo {year} {2020})}\BibitemShut {NoStop}%
\bibitem [{\citenamefont {Yuan}\ \emph {et~al.}(2020)\citenamefont {Yuan}, \citenamefont {Wang}, \citenamefont {Luo}, \citenamefont {Rashba},\ and\ \citenamefont {Zunger}}]{Yuan20}%
  \BibitemOpen
  \bibfield  {author} {\bibinfo {author} {\bibfnamefont {L.-D.}\ \bibnamefont {Yuan}}, \bibinfo {author} {\bibfnamefont {Z.}~\bibnamefont {Wang}}, \bibinfo {author} {\bibfnamefont {J.-W.}\ \bibnamefont {Luo}}, \bibinfo {author} {\bibfnamefont {E.~I.}\ \bibnamefont {Rashba}},\ and\ \bibinfo {author} {\bibfnamefont {A.}~\bibnamefont {Zunger}},\ }\bibfield  {title} {\bibinfo {title} {Giant momentum-dependent spin splitting in centrosymmetric {low-$Z$} antiferromagnets},\ }\href {https://doi.org/10.1103/PhysRevB.102.014422} {\bibfield  {journal} {\bibinfo  {journal} {Phys. Rev. B}\ }\textbf {\bibinfo {volume} {102}},\ \bibinfo {pages} {014422} (\bibinfo {year} {2020})}\BibitemShut {NoStop}%
\bibitem [{\citenamefont {Yuan}\ \emph {et~al.}(2021)\citenamefont {Yuan}, \citenamefont {Wang}, \citenamefont {Luo},\ and\ \citenamefont {Zunger}}]{Yuan21}%
  \BibitemOpen
  \bibfield  {author} {\bibinfo {author} {\bibfnamefont {L.-D.}\ \bibnamefont {Yuan}}, \bibinfo {author} {\bibfnamefont {Z.}~\bibnamefont {Wang}}, \bibinfo {author} {\bibfnamefont {J.-W.}\ \bibnamefont {Luo}},\ and\ \bibinfo {author} {\bibfnamefont {A.}~\bibnamefont {Zunger}},\ }\bibfield  {title} {\bibinfo {title} {Prediction of {low-Z} collinear and noncollinear antiferromagnetic compounds having momentum-dependent spin splitting even without spin-orbit coupling},\ }\href {https://doi.org/10.1103/PhysRevMaterials.5.014409} {\bibfield  {journal} {\bibinfo  {journal} {Phys. Rev. Mater.}\ }\textbf {\bibinfo {volume} {5}},\ \bibinfo {pages} {014409} (\bibinfo {year} {2021})}\BibitemShut {NoStop}%
\bibitem [{\citenamefont {Hayami}\ \emph {et~al.}(2020)\citenamefont {Hayami}, \citenamefont {Yanagi},\ and\ \citenamefont {Kusunose}}]{Hayami20}%
  \BibitemOpen
  \bibfield  {author} {\bibinfo {author} {\bibfnamefont {S.}~\bibnamefont {Hayami}}, \bibinfo {author} {\bibfnamefont {Y.}~\bibnamefont {Yanagi}},\ and\ \bibinfo {author} {\bibfnamefont {H.}~\bibnamefont {Kusunose}},\ }\bibfield  {title} {\bibinfo {title} {Bottom-up design of spin-split and reshaped electronic band structures in antiferromagnets without spin-orbit coupling: Procedure on the basis of augmented multipoles},\ }\href {https://doi.org/10.1103/PhysRevB.102.144441} {\bibfield  {journal} {\bibinfo  {journal} {Phys. Rev. B}\ }\textbf {\bibinfo {volume} {102}},\ \bibinfo {pages} {144441} (\bibinfo {year} {2020})}\BibitemShut {NoStop}%
\bibitem [{\citenamefont {Mazin}\ \emph {et~al.}(2021)\citenamefont {Mazin}, \citenamefont {Koepernik}, \citenamefont {Johannes}, \citenamefont {González-Hernández},\ and\ \citenamefont {\v{S}mejkal}}]{Mazin21}%
  \BibitemOpen
  \bibfield  {author} {\bibinfo {author} {\bibfnamefont {I.~I.}\ \bibnamefont {Mazin}}, \bibinfo {author} {\bibfnamefont {K.}~\bibnamefont {Koepernik}}, \bibinfo {author} {\bibfnamefont {M.~D.}\ \bibnamefont {Johannes}}, \bibinfo {author} {\bibfnamefont {R.}~\bibnamefont {González-Hernández}},\ and\ \bibinfo {author} {\bibfnamefont {L.}~\bibnamefont {\v{S}mejkal}},\ }\bibfield  {title} {\bibinfo {title} {Prediction of unconventional magnetism in doped {FeSb$_2$}},\ }\href {https://doi.org/10.1073/pnas.2108924118} {\bibfield  {journal} {\bibinfo  {journal} {Proc. Natl. Acad. Sci. U.S.A.}\ }\textbf {\bibinfo {volume} {118}},\ \bibinfo {pages} {e2108924118} (\bibinfo {year} {2021})}\BibitemShut {NoStop}%
\bibitem [{\citenamefont {Liu}\ \emph {et~al.}(2022)\citenamefont {Liu}, \citenamefont {Li}, \citenamefont {Han}, \citenamefont {Wan},\ and\ \citenamefont {Liu}}]{Liu22}%
  \BibitemOpen
  \bibfield  {author} {\bibinfo {author} {\bibfnamefont {P.}~\bibnamefont {Liu}}, \bibinfo {author} {\bibfnamefont {J.}~\bibnamefont {Li}}, \bibinfo {author} {\bibfnamefont {J.}~\bibnamefont {Han}}, \bibinfo {author} {\bibfnamefont {X.}~\bibnamefont {Wan}},\ and\ \bibinfo {author} {\bibfnamefont {Q.}~\bibnamefont {Liu}},\ }\bibfield  {title} {\bibinfo {title} {Spin-{G}roup {S}ymmetry in {M}agnetic {M}aterials with {N}egligible {S}pin-{O}rbit {C}oupling},\ }\href {https://doi.org/10.1103/PhysRevX.12.021016} {\bibfield  {journal} {\bibinfo  {journal} {Phys. Rev. X}\ }\textbf {\bibinfo {volume} {12}},\ \bibinfo {pages} {021016} (\bibinfo {year} {2022})}\BibitemShut {NoStop}%
\bibitem [{\citenamefont {Yang}\ \emph {et~al.}()\citenamefont {Yang}, \citenamefont {Liu},\ and\ \citenamefont {Fang}}]{Jian23}%
  \BibitemOpen
  \bibfield  {author} {\bibinfo {author} {\bibfnamefont {J.}~\bibnamefont {Yang}}, \bibinfo {author} {\bibfnamefont {Z.-X.}\ \bibnamefont {Liu}},\ and\ \bibinfo {author} {\bibfnamefont {C.}~\bibnamefont {Fang}},\ }\href@noop {} {\bibinfo {title} {Symmetry invariants and classes of quasi-particles in magnetically ordered systems having weak spin-orbit coupling}},\ \Eprint {https://arxiv.org/abs/2105.12738} {arXiv:2105.12738} \BibitemShut {NoStop}%
\bibitem [{\citenamefont {{\v{S}}mejkal}\ \emph {et~al.}(2022)\citenamefont {{\v{S}}mejkal}, \citenamefont {MacDonald}, \citenamefont {Sinova}, \citenamefont {Nakatsuji},\ and\ \citenamefont {Jungwirth}}]{Smejkal22b}%
  \BibitemOpen
  \bibfield  {author} {\bibinfo {author} {\bibfnamefont {L.}~\bibnamefont {{\v{S}}mejkal}}, \bibinfo {author} {\bibfnamefont {A.~H.}\ \bibnamefont {MacDonald}}, \bibinfo {author} {\bibfnamefont {J.}~\bibnamefont {Sinova}}, \bibinfo {author} {\bibfnamefont {S.}~\bibnamefont {Nakatsuji}},\ and\ \bibinfo {author} {\bibfnamefont {T.}~\bibnamefont {Jungwirth}},\ }\bibfield  {title} {\bibinfo {title} {Anomalous {Hall} antiferromagnets},\ }\href {https://doi.org/10.1038/s41578-022-00430-3} {\bibfield  {journal} {\bibinfo  {journal} {Nat. Rev. Mater.}\ }\textbf {\bibinfo {volume} {7}},\ \bibinfo {pages} {482} (\bibinfo {year} {2022})}\BibitemShut {NoStop}%
\bibitem [{\citenamefont {Samanta}\ \emph {et~al.}(2020)\citenamefont {Samanta}, \citenamefont {Ležaić}, \citenamefont {Merte}, \citenamefont {Freimuth}, \citenamefont {Blügel},\ and\ \citenamefont {Mokrousov}}]{Samanta20}%
  \BibitemOpen
  \bibfield  {author} {\bibinfo {author} {\bibfnamefont {K.}~\bibnamefont {Samanta}}, \bibinfo {author} {\bibfnamefont {M.}~\bibnamefont {Ležaić}}, \bibinfo {author} {\bibfnamefont {M.}~\bibnamefont {Merte}}, \bibinfo {author} {\bibfnamefont {F.}~\bibnamefont {Freimuth}}, \bibinfo {author} {\bibfnamefont {S.}~\bibnamefont {Blügel}},\ and\ \bibinfo {author} {\bibfnamefont {Y.}~\bibnamefont {Mokrousov}},\ }\bibfield  {title} {\bibinfo {title} {{Crystal {Hall} and crystal magneto-optical effect in thin films of SrRuO{$_3$}}},\ }\href {https://doi.org/10.1063/5.0005017} {\bibfield  {journal} {\bibinfo  {journal} {J. Appl. Phys.}\ }\textbf {\bibinfo {volume} {127}},\ \bibinfo {pages} {213904} (\bibinfo {year} {2020})}\BibitemShut {NoStop}%
\bibitem [{\citenamefont {Naka}\ \emph {et~al.}(2020)\citenamefont {Naka}, \citenamefont {Hayami}, \citenamefont {Kusunose}, \citenamefont {Yanagi}, \citenamefont {Motome},\ and\ \citenamefont {Seo}}]{Naka20}%
  \BibitemOpen
  \bibfield  {author} {\bibinfo {author} {\bibfnamefont {M.}~\bibnamefont {Naka}}, \bibinfo {author} {\bibfnamefont {S.}~\bibnamefont {Hayami}}, \bibinfo {author} {\bibfnamefont {H.}~\bibnamefont {Kusunose}}, \bibinfo {author} {\bibfnamefont {Y.}~\bibnamefont {Yanagi}}, \bibinfo {author} {\bibfnamefont {Y.}~\bibnamefont {Motome}},\ and\ \bibinfo {author} {\bibfnamefont {H.}~\bibnamefont {Seo}},\ }\bibfield  {title} {\bibinfo {title} {Anomalous {Hall} effect in $\ensuremath{\kappa}$-type organic antiferromagnets},\ }\href {https://doi.org/10.1103/PhysRevB.102.075112} {\bibfield  {journal} {\bibinfo  {journal} {Phys. Rev. B}\ }\textbf {\bibinfo {volume} {102}},\ \bibinfo {pages} {075112} (\bibinfo {year} {2020})}\BibitemShut {NoStop}%
\bibitem [{\citenamefont {Hayami}\ and\ \citenamefont {Kusunose}(2021)}]{Hayami21}%
  \BibitemOpen
  \bibfield  {author} {\bibinfo {author} {\bibfnamefont {S.}~\bibnamefont {Hayami}}\ and\ \bibinfo {author} {\bibfnamefont {H.}~\bibnamefont {Kusunose}},\ }\bibfield  {title} {\bibinfo {title} {Essential role of the anisotropic magnetic dipole in the anomalous {Hall} effect},\ }\href {https://doi.org/10.1103/PhysRevB.103.L180407} {\bibfield  {journal} {\bibinfo  {journal} {Phys. Rev. B}\ }\textbf {\bibinfo {volume} {103}},\ \bibinfo {pages} {L180407} (\bibinfo {year} {2021})}\BibitemShut {NoStop}%
\bibitem [{\citenamefont {Gonzalez~Betancourt}\ \emph {et~al.}(2023)\citenamefont {Gonzalez~Betancourt}, \citenamefont {Zub\'a\ifmmode~\check{c}\else \v{c}\fi{}}, \citenamefont {Gonzalez-Hernandez}, \citenamefont {Geishendorf}, \citenamefont {\ifmmode \check{S}\else \v{S}\fi{}ob\'a\ifmmode~\check{n}\else \v{n}\fi{}}, \citenamefont {Springholz}, \citenamefont {Olejn\'{\i}k}, \citenamefont {\ifmmode~\check{S}\else \v{S}\fi{}mejkal}, \citenamefont {Sinova}, \citenamefont {Jungwirth}, \citenamefont {Goennenwein}, \citenamefont {Thomas}, \citenamefont {Reichlov\'a}, \citenamefont {\ifmmode~\check{Z}\else \v{Z}\fi{}elezn\'y},\ and\ \citenamefont {Kriegner}}]{Gonzalez2023}%
  \BibitemOpen
  \bibfield  {author} {\bibinfo {author} {\bibfnamefont {R.~D.}\ \bibnamefont {Gonzalez~Betancourt}}, \bibinfo {author} {\bibfnamefont {J.}~\bibnamefont {Zub\'a\ifmmode~\check{c}\else \v{c}\fi{}}}, \bibinfo {author} {\bibfnamefont {R.}~\bibnamefont {Gonzalez-Hernandez}}, \bibinfo {author} {\bibfnamefont {K.}~\bibnamefont {Geishendorf}}, \bibinfo {author} {\bibfnamefont {Z.}~\bibnamefont {\ifmmode \check{S}\else \v{S}\fi{}ob\'a\ifmmode~\check{n}\else \v{n}\fi{}}}, \bibinfo {author} {\bibfnamefont {G.}~\bibnamefont {Springholz}}, \bibinfo {author} {\bibfnamefont {K.}~\bibnamefont {Olejn\'{\i}k}}, \bibinfo {author} {\bibfnamefont {L.}~\bibnamefont {\ifmmode~\check{S}\else \v{S}\fi{}mejkal}}, \bibinfo {author} {\bibfnamefont {J.}~\bibnamefont {Sinova}}, \bibinfo {author} {\bibfnamefont {T.}~\bibnamefont {Jungwirth}}, \bibinfo {author} {\bibfnamefont {S.~T.~B.}\ \bibnamefont {Goennenwein}}, \bibinfo {author} {\bibfnamefont {A.}~\bibnamefont {Thomas}}, \bibinfo {author} {\bibfnamefont {H.}~\bibnamefont
  {Reichlov\'a}}, \bibinfo {author} {\bibfnamefont {J.}~\bibnamefont {\ifmmode~\check{Z}\else \v{Z}\fi{}elezn\'y}},\ and\ \bibinfo {author} {\bibfnamefont {D.}~\bibnamefont {Kriegner}},\ }\bibfield  {title} {\bibinfo {title} {Spontaneous {A}nomalous {Hall} {E}ffect {A}rising from an {U}nconventional {C}ompensated {M}agnetic {P}hase in a {S}emiconductor},\ }\href {https://doi.org/10.1103/PhysRevLett.130.036702} {\bibfield  {journal} {\bibinfo  {journal} {Phys. Rev. Lett.}\ }\textbf {\bibinfo {volume} {130}},\ \bibinfo {pages} {036702} (\bibinfo {year} {2023})}\BibitemShut {NoStop}%
\bibitem [{\citenamefont {Naka}\ \emph {et~al.}(2022)\citenamefont {Naka}, \citenamefont {Motome},\ and\ \citenamefont {Seo}}]{Naka22}%
  \BibitemOpen
  \bibfield  {author} {\bibinfo {author} {\bibfnamefont {M.}~\bibnamefont {Naka}}, \bibinfo {author} {\bibfnamefont {Y.}~\bibnamefont {Motome}},\ and\ \bibinfo {author} {\bibfnamefont {H.}~\bibnamefont {Seo}},\ }\bibfield  {title} {\bibinfo {title} {Anomalous {Hall} effect in antiferromagnetic perovskites},\ }\href {https://doi.org/10.1103/PhysRevB.106.195149} {\bibfield  {journal} {\bibinfo  {journal} {Phys. Rev. B}\ }\textbf {\bibinfo {volume} {106}},\ \bibinfo {pages} {195149} (\bibinfo {year} {2022})}\BibitemShut {NoStop}%
\bibitem [{\citenamefont {Hariki}\ \emph {et~al.}(2024{\natexlab{a}})\citenamefont {Hariki}, \citenamefont {Takahashi},\ and\ \citenamefont {Kune\ifmmode~\check{s}\else \v{s}\fi{}}}]{Hariki2024a}%
  \BibitemOpen
  \bibfield  {author} {\bibinfo {author} {\bibfnamefont {A.}~\bibnamefont {Hariki}}, \bibinfo {author} {\bibfnamefont {Y.}~\bibnamefont {Takahashi}},\ and\ \bibinfo {author} {\bibfnamefont {J.}~\bibnamefont {Kune\ifmmode~\check{s}\else \v{s}\fi{}}},\ }\bibfield  {title} {\bibinfo {title} {X-ray magnetic circular dichroism in {R}u{O}$_{2}$},\ }\href {https://doi.org/10.1103/PhysRevB.109.094413} {\bibfield  {journal} {\bibinfo  {journal} {Phys. Rev. B}\ }\textbf {\bibinfo {volume} {109}},\ \bibinfo {pages} {094413} (\bibinfo {year} {2024}{\natexlab{a}})}\BibitemShut {NoStop}%
\bibitem [{\citenamefont {Hariki}\ \emph {et~al.}(2024{\natexlab{b}})\citenamefont {Hariki}, \citenamefont {Dal~Din}, \citenamefont {Amin}, \citenamefont {Yamaguchi}, \citenamefont {Badura}, \citenamefont {Kriegner}, \citenamefont {Edmonds}, \citenamefont {Campion}, \citenamefont {Wadley}, \citenamefont {Backes}, \citenamefont {Veiga}, \citenamefont {Dhesi}, \citenamefont {Springholz}, \citenamefont {\ifmmode~\check{S}\else \v{S}\fi{}mejkal}, \citenamefont {V\'yborn\'y}, \citenamefont {Jungwirth},\ and\ \citenamefont {Kune\ifmmode~\check{s}\else \v{s}\fi{}}}]{Hariki2024b}%
  \BibitemOpen
  \bibfield  {author} {\bibinfo {author} {\bibfnamefont {A.}~\bibnamefont {Hariki}}, \bibinfo {author} {\bibfnamefont {A.}~\bibnamefont {Dal~Din}}, \bibinfo {author} {\bibfnamefont {O.~J.}\ \bibnamefont {Amin}}, \bibinfo {author} {\bibfnamefont {T.}~\bibnamefont {Yamaguchi}}, \bibinfo {author} {\bibfnamefont {A.}~\bibnamefont {Badura}}, \bibinfo {author} {\bibfnamefont {D.}~\bibnamefont {Kriegner}}, \bibinfo {author} {\bibfnamefont {K.~W.}\ \bibnamefont {Edmonds}}, \bibinfo {author} {\bibfnamefont {R.~P.}\ \bibnamefont {Campion}}, \bibinfo {author} {\bibfnamefont {P.}~\bibnamefont {Wadley}}, \bibinfo {author} {\bibfnamefont {D.}~\bibnamefont {Backes}}, \bibinfo {author} {\bibfnamefont {L.~S.~I.}\ \bibnamefont {Veiga}}, \bibinfo {author} {\bibfnamefont {S.~S.}\ \bibnamefont {Dhesi}}, \bibinfo {author} {\bibfnamefont {G.}~\bibnamefont {Springholz}}, \bibinfo {author} {\bibfnamefont {L.}~\bibnamefont {\ifmmode~\check{S}\else \v{S}\fi{}mejkal}}, \bibinfo {author} {\bibfnamefont {K.}~\bibnamefont {V\'yborn\'y}},
  \bibinfo {author} {\bibfnamefont {T.}~\bibnamefont {Jungwirth}},\ and\ \bibinfo {author} {\bibfnamefont {J.}~\bibnamefont {Kune\ifmmode~\check{s}\else \v{s}\fi{}}},\ }\bibfield  {title} {\bibinfo {title} {X-ray {M}agnetic {C}ircular {D}ichroism in {A}ltermagnetic $\ensuremath{\alpha}$-{M}n{T}e},\ }\href {https://doi.org/10.1103/PhysRevLett.132.176701} {\bibfield  {journal} {\bibinfo  {journal} {Phys. Rev. Lett.}\ }\textbf {\bibinfo {volume} {132}},\ \bibinfo {pages} {176701} (\bibinfo {year} {2024}{\natexlab{b}})}\BibitemShut {NoStop}%
\bibitem [{\citenamefont {Sasabe}\ \emph {et~al.}(2023)\citenamefont {Sasabe}, \citenamefont {Mizumaki}, \citenamefont {Uozumi},\ and\ \citenamefont {Yamasaki}}]{Sasabe23}%
  \BibitemOpen
  \bibfield  {author} {\bibinfo {author} {\bibfnamefont {N.}~\bibnamefont {Sasabe}}, \bibinfo {author} {\bibfnamefont {M.}~\bibnamefont {Mizumaki}}, \bibinfo {author} {\bibfnamefont {T.}~\bibnamefont {Uozumi}},\ and\ \bibinfo {author} {\bibfnamefont {Y.}~\bibnamefont {Yamasaki}},\ }\bibfield  {title} {\bibinfo {title} {Ferroic {O}rder for {A}nisotropic {M}agnetic {D}ipole {T}erm in {C}ollinear {A}ntiferromagnets of $({t}_{2g}{)}^{4}$ {S}ystem},\ }\href {https://doi.org/10.1103/PhysRevLett.131.216501} {\bibfield  {journal} {\bibinfo  {journal} {Phys. Rev. Lett.}\ }\textbf {\bibinfo {volume} {131}},\ \bibinfo {pages} {216501} (\bibinfo {year} {2023})}\BibitemShut {NoStop}%
\bibitem [{\citenamefont {Watanabe}\ \emph {et~al.}(2024)\citenamefont {Watanabe}, \citenamefont {Shinohara}, \citenamefont {Nomoto}, \citenamefont {Togo},\ and\ \citenamefont {Arita}}]{Watanabe2024}%
  \BibitemOpen
  \bibfield  {author} {\bibinfo {author} {\bibfnamefont {H.}~\bibnamefont {Watanabe}}, \bibinfo {author} {\bibfnamefont {K.}~\bibnamefont {Shinohara}}, \bibinfo {author} {\bibfnamefont {T.}~\bibnamefont {Nomoto}}, \bibinfo {author} {\bibfnamefont {A.}~\bibnamefont {Togo}},\ and\ \bibinfo {author} {\bibfnamefont {R.}~\bibnamefont {Arita}},\ }\bibfield  {title} {\bibinfo {title} {Symmetry analysis with spin crystallographic groups: Disentangling effects free of spin-orbit coupling in emergent electromagnetism},\ }\href {https://doi.org/10.1103/PhysRevB.109.094438} {\bibfield  {journal} {\bibinfo  {journal} {Phys. Rev. B}\ }\textbf {\bibinfo {volume} {109}},\ \bibinfo {pages} {094438} (\bibinfo {year} {2024})}\BibitemShut {NoStop}%
\bibitem [{\citenamefont {Feng}\ \emph {et~al.}(2022)\citenamefont {Feng}, \citenamefont {Zhou}, \citenamefont {{\v{S}}mejkal}, \citenamefont {Wu}, \citenamefont {Zhu}, \citenamefont {Guo}, \citenamefont {Gonz{\'a}lez-Hern{\'a}ndez}, \citenamefont {Wang}, \citenamefont {Yan}, \citenamefont {Qin}, \citenamefont {Zhang}, \citenamefont {Wu}, \citenamefont {Chen}, \citenamefont {Meng}, \citenamefont {Liu}, \citenamefont {Xia}, \citenamefont {Sinova}, \citenamefont {Jungwirth},\ and\ \citenamefont {Liu}}]{Feng2022}%
  \BibitemOpen
  \bibfield  {author} {\bibinfo {author} {\bibfnamefont {Z.}~\bibnamefont {Feng}}, \bibinfo {author} {\bibfnamefont {X.}~\bibnamefont {Zhou}}, \bibinfo {author} {\bibfnamefont {L.}~\bibnamefont {{\v{S}}mejkal}}, \bibinfo {author} {\bibfnamefont {L.}~\bibnamefont {Wu}}, \bibinfo {author} {\bibfnamefont {Z.}~\bibnamefont {Zhu}}, \bibinfo {author} {\bibfnamefont {H.}~\bibnamefont {Guo}}, \bibinfo {author} {\bibfnamefont {R.}~\bibnamefont {Gonz{\'a}lez-Hern{\'a}ndez}}, \bibinfo {author} {\bibfnamefont {X.}~\bibnamefont {Wang}}, \bibinfo {author} {\bibfnamefont {H.}~\bibnamefont {Yan}}, \bibinfo {author} {\bibfnamefont {P.}~\bibnamefont {Qin}}, \bibinfo {author} {\bibfnamefont {X.}~\bibnamefont {Zhang}}, \bibinfo {author} {\bibfnamefont {H.}~\bibnamefont {Wu}}, \bibinfo {author} {\bibfnamefont {H.}~\bibnamefont {Chen}}, \bibinfo {author} {\bibfnamefont {Z.}~\bibnamefont {Meng}}, \bibinfo {author} {\bibfnamefont {L.}~\bibnamefont {Liu}}, \bibinfo {author} {\bibfnamefont {Z.}~\bibnamefont {Xia}}, \bibinfo {author}
  {\bibfnamefont {J.}~\bibnamefont {Sinova}}, \bibinfo {author} {\bibfnamefont {T.}~\bibnamefont {Jungwirth}},\ and\ \bibinfo {author} {\bibfnamefont {Z.}~\bibnamefont {Liu}},\ }\bibfield  {title} {\bibinfo {title} {An anomalous {Hall} effect in altermagnetic ruthenium dioxide},\ }\href {https://doi.org/10.1038/s41928-022-00866-z} {\bibfield  {journal} {\bibinfo  {journal} {Nat. Electron.}\ }\textbf {\bibinfo {volume} {5}},\ \bibinfo {pages} {735} (\bibinfo {year} {2022})}\BibitemShut {NoStop}%
\bibitem [{\citenamefont {Jim{\'e}nez-Mier}\ \emph {et~al.}(2017)\citenamefont {Jim{\'e}nez-Mier}, \citenamefont {Olalde-Velasco}, \citenamefont {de~la Mora}, \citenamefont {Yang},\ and\ \citenamefont {Denlinger}}]{Jose17}%
  \BibitemOpen
  \bibfield  {author} {\bibinfo {author} {\bibfnamefont {J.}~\bibnamefont {Jim{\'e}nez-Mier}}, \bibinfo {author} {\bibfnamefont {P.}~\bibnamefont {Olalde-Velasco}}, \bibinfo {author} {\bibfnamefont {P.}~\bibnamefont {de~la Mora}}, \bibinfo {author} {\bibfnamefont {W.}~\bibnamefont {Yang}},\ and\ \bibinfo {author} {\bibfnamefont {J.~D.}\ \bibnamefont {Denlinger}},\ }\bibfield  {title} {\bibinfo {title} {Atomic multiplet and charge transfer effects in the resonant inelastic x-ray scattering (rixs) spectra at the nickel l2,3 edge of nif2},\ }\href {https://api.semanticscholar.org/CorpusID:52260138} {\bibfield  {journal} {\bibinfo  {journal} {Journal of Nuclear Physics, Material Sciences, Radiation and Applications}\ }\textbf {\bibinfo {volume} {5}},\ \bibinfo {pages} {1} (\bibinfo {year} {2017})}\BibitemShut {NoStop}%
\bibitem [{\citenamefont {Kune\ifmmode~\check{s}\else \v{s}\fi{}}\ and\ \citenamefont {Oppeneer}(2003)}]{Kunes2003}%
  \BibitemOpen
  \bibfield  {author} {\bibinfo {author} {\bibfnamefont {J.}~\bibnamefont {Kune\ifmmode~\check{s}\else \v{s}\fi{}}}\ and\ \bibinfo {author} {\bibfnamefont {P.~M.}\ \bibnamefont {Oppeneer}},\ }\bibfield  {title} {\bibinfo {title} {Anisotropic {X}-ray magnetic linear dichroism at the ${L}_{2,3}$ edges of cubic {Fe, Co, and Ni}: $\mathit{Ab}$ $\mathit{initio}$ calculations and model theory},\ }\href {https://doi.org/10.1103/PhysRevB.67.024431} {\bibfield  {journal} {\bibinfo  {journal} {Phys. Rev. B}\ }\textbf {\bibinfo {volume} {67}},\ \bibinfo {pages} {024431} (\bibinfo {year} {2003})}\BibitemShut {NoStop}%
\bibitem [{\citenamefont {Stout}\ and\ \citenamefont {Catalano}(1953)}]{Stout1953}%
  \BibitemOpen
  \bibfield  {author} {\bibinfo {author} {\bibfnamefont {J.~W.}\ \bibnamefont {Stout}}\ and\ \bibinfo {author} {\bibfnamefont {E.}~\bibnamefont {Catalano}},\ }\bibfield  {title} {\bibinfo {title} {Thermal anomalies associated with the antiferromagnetic ordering of {Fe${\mathrm{F}}_{2}$}, co${\mathrm{f}}_{2}$, and ni${\mathrm{f}}_{2}$},\ }\href {https://doi.org/10.1103/PhysRev.92.1575} {\bibfield  {journal} {\bibinfo  {journal} {Phys. Rev.}\ }\textbf {\bibinfo {volume} {92}},\ \bibinfo {pages} {1575} (\bibinfo {year} {1953})}\BibitemShut {NoStop}%
\bibitem [{\citenamefont {Erickson}(1953)}]{Erickson1953}%
  \BibitemOpen
  \bibfield  {author} {\bibinfo {author} {\bibfnamefont {R.~A.}\ \bibnamefont {Erickson}},\ }\bibfield  {title} {\bibinfo {title} {Neutron {D}iffraction {S}tudies of {A}ntiferromagnetism in {M}anganous {F}luoride and {S}ome {I}somorphous {C}ompounds},\ }\href {https://doi.org/10.1103/PhysRev.90.779} {\bibfield  {journal} {\bibinfo  {journal} {Phys. Rev.}\ }\textbf {\bibinfo {volume} {90}},\ \bibinfo {pages} {779} (\bibinfo {year} {1953})}\BibitemShut {NoStop}%
\bibitem [{\citenamefont {Matarrese}\ and\ \citenamefont {Stout}(1954)}]{Matarrese1954}%
  \BibitemOpen
  \bibfield  {author} {\bibinfo {author} {\bibfnamefont {L.~M.}\ \bibnamefont {Matarrese}}\ and\ \bibinfo {author} {\bibfnamefont {J.~W.}\ \bibnamefont {Stout}},\ }\bibfield  {title} {\bibinfo {title} {Magnetic anisotropy of {Ni${\mathrm{F}}_{2}$}},\ }\href {https://doi.org/10.1103/PhysRev.94.1792} {\bibfield  {journal} {\bibinfo  {journal} {Phys. Rev.}\ }\textbf {\bibinfo {volume} {94}},\ \bibinfo {pages} {1792} (\bibinfo {year} {1954})}\BibitemShut {NoStop}%
\bibitem [{\citenamefont {Hariki}\ \emph {et~al.}(2024{\natexlab{c}})\citenamefont {Hariki}, \citenamefont {Okauchi}, \citenamefont {Takahashi},\ and\ \citenamefont {Kune\ifmmode~\check{s}\else \v{s}\fi{}}}]{MnF2}%
  \BibitemOpen
  \bibfield  {author} {\bibinfo {author} {\bibfnamefont {A.}~\bibnamefont {Hariki}}, \bibinfo {author} {\bibfnamefont {T.}~\bibnamefont {Okauchi}}, \bibinfo {author} {\bibfnamefont {Y.}~\bibnamefont {Takahashi}},\ and\ \bibinfo {author} {\bibfnamefont {J.}~\bibnamefont {Kune\ifmmode~\check{s}\else \v{s}\fi{}}},\ }\bibfield  {title} {\bibinfo {title} {Determination of the n\'eel vector in rutile altermagnets through x-ray magnetic circular dichroism: The case of {${\mathrm{MnF}}_{2}$}},\ }\href {https://doi.org/10.1103/PhysRevB.110.L100402} {\bibfield  {journal} {\bibinfo  {journal} {Phys. Rev. B}\ }\textbf {\bibinfo {volume} {110}},\ \bibinfo {pages} {L100402} (\bibinfo {year} {2024}{\natexlab{c}})}\BibitemShut {NoStop}%
\bibitem [{\citenamefont {Stoji\ifmmode~\acute{c}\else \'{c}\fi{}}\ \emph {et~al.}(2005)\citenamefont {Stoji\ifmmode~\acute{c}\else \'{c}\fi{}}, \citenamefont {Binggeli},\ and\ \citenamefont {Altarelli}}]{Stojic2005}%
  \BibitemOpen
  \bibfield  {author} {\bibinfo {author} {\bibfnamefont {N.}~\bibnamefont {Stoji\ifmmode~\acute{c}\else \'{c}\fi{}}}, \bibinfo {author} {\bibfnamefont {N.}~\bibnamefont {Binggeli}},\ and\ \bibinfo {author} {\bibfnamefont {M.}~\bibnamefont {Altarelli}},\ }\bibfield  {title} {\bibinfo {title} {$\mathrm{Mn}\phantom{\rule{0.2em}{0ex}}{L}_{2,3}$ edge resonant x-ray scattering in manganites: Influence of the magnetic state},\ }\href {https://doi.org/10.1103/PhysRevB.72.104108} {\bibfield  {journal} {\bibinfo  {journal} {Phys. Rev. B}\ }\textbf {\bibinfo {volume} {72}},\ \bibinfo {pages} {104108} (\bibinfo {year} {2005})}\BibitemShut {NoStop}%
\bibitem [{\citenamefont {Arenholz}\ \emph {et~al.}(2006)\citenamefont {Arenholz}, \citenamefont {van~der Laan}, \citenamefont {Chopdekar},\ and\ \citenamefont {Suzuki}}]{Arenholz2006}%
  \BibitemOpen
  \bibfield  {author} {\bibinfo {author} {\bibfnamefont {E.}~\bibnamefont {Arenholz}}, \bibinfo {author} {\bibfnamefont {G.}~\bibnamefont {van~der Laan}}, \bibinfo {author} {\bibfnamefont {R.~V.}\ \bibnamefont {Chopdekar}},\ and\ \bibinfo {author} {\bibfnamefont {Y.}~\bibnamefont {Suzuki}},\ }\bibfield  {title} {\bibinfo {title} {Anisotropic x-ray magnetic linear dichroism at the fe ${L}_{2,3}$ edges in ${\mathrm{fe}}_{3}{\mathrm{o}}_{4}$},\ }\href {https://doi.org/10.1103/PhysRevB.74.094407} {\bibfield  {journal} {\bibinfo  {journal} {Phys. Rev. B}\ }\textbf {\bibinfo {volume} {74}},\ \bibinfo {pages} {094407} (\bibinfo {year} {2006})}\BibitemShut {NoStop}%
\bibitem [{\citenamefont {Arenholz}\ \emph {et~al.}(2007)\citenamefont {Arenholz}, \citenamefont {van~der Laan}, \citenamefont {Chopdekar},\ and\ \citenamefont {Suzuki}}]{Arenholz2007}%
  \BibitemOpen
  \bibfield  {author} {\bibinfo {author} {\bibfnamefont {E.}~\bibnamefont {Arenholz}}, \bibinfo {author} {\bibfnamefont {G.}~\bibnamefont {van~der Laan}}, \bibinfo {author} {\bibfnamefont {R.~V.}\ \bibnamefont {Chopdekar}},\ and\ \bibinfo {author} {\bibfnamefont {Y.}~\bibnamefont {Suzuki}},\ }\bibfield  {title} {\bibinfo {title} {Angle-dependent ${\mathrm{ni}}^{2+}$ x-ray magnetic linear dichroism: Interfacial coupling revisited},\ }\href {https://doi.org/10.1103/PhysRevLett.98.197201} {\bibfield  {journal} {\bibinfo  {journal} {Phys. Rev. Lett.}\ }\textbf {\bibinfo {volume} {98}},\ \bibinfo {pages} {197201} (\bibinfo {year} {2007})}\BibitemShut {NoStop}%
\bibitem [{\citenamefont {Haverkort}\ \emph {et~al.}(2010)\citenamefont {Haverkort}, \citenamefont {Hollmann}, \citenamefont {Krug},\ and\ \citenamefont {Tanaka}}]{Haverkort2010}%
  \BibitemOpen
  \bibfield  {author} {\bibinfo {author} {\bibfnamefont {M.~W.}\ \bibnamefont {Haverkort}}, \bibinfo {author} {\bibfnamefont {N.}~\bibnamefont {Hollmann}}, \bibinfo {author} {\bibfnamefont {I.~P.}\ \bibnamefont {Krug}},\ and\ \bibinfo {author} {\bibfnamefont {A.}~\bibnamefont {Tanaka}},\ }\bibfield  {title} {\bibinfo {title} {Symmetry analysis of magneto-optical effects: The case of x-ray diffraction and x-ray absorption at the transition metal ${L}_{2,3}$ edge},\ }\href {https://doi.org/10.1103/PhysRevB.82.094403} {\bibfield  {journal} {\bibinfo  {journal} {Phys. Rev. B}\ }\textbf {\bibinfo {volume} {82}},\ \bibinfo {pages} {094403} (\bibinfo {year} {2010})}\BibitemShut {NoStop}%
\bibitem [{sm()}]{sm}%
  \BibitemOpen
  \href@noop {} {}\bibinfo {note} {See Supplementary Material for details at ...}\BibitemShut {Stop}%
\bibitem [{\citenamefont {Borovik-Romanov}\ \emph {et~al.}(1973)\citenamefont {Borovik-Romanov}, \citenamefont {Bazhan},\ and\ \citenamefont {Kreines}}]{Borovik1973}%
  \BibitemOpen
  \bibfield  {author} {\bibinfo {author} {\bibfnamefont {A.~S.}\ \bibnamefont {Borovik-Romanov}}, \bibinfo {author} {\bibfnamefont {A.~N.}\ \bibnamefont {Bazhan}},\ and\ \bibinfo {author} {\bibfnamefont {N.~M.}\ \bibnamefont {Kreines}},\ }\bibfield  {title} {\bibinfo {title} {The weak ferromagnetism of {NiF$_{2}$}},\ }\href@noop {} {\bibfield  {journal} {\bibinfo  {journal} {Zh. Eksp. Teor. Fiz.}\ }\textbf {\bibinfo {volume} {64}},\ \bibinfo {pages} {1367} (\bibinfo {year} {1973})}\BibitemShut {NoStop}%
\bibitem [{\citenamefont {Moriya}(1960)}]{Moriya1960}%
  \BibitemOpen
  \bibfield  {author} {\bibinfo {author} {\bibfnamefont {T.}~\bibnamefont {Moriya}},\ }\bibfield  {title} {\bibinfo {title} {Theory of magnetism of {Ni${\mathrm{F}}_{2}$}},\ }\href {https://doi.org/10.1103/PhysRev.117.635} {\bibfield  {journal} {\bibinfo  {journal} {Phys. Rev.}\ }\textbf {\bibinfo {volume} {117}},\ \bibinfo {pages} {635} (\bibinfo {year} {1960})}\BibitemShut {NoStop}%
\bibitem [{\citenamefont {Thole}\ \emph {et~al.}(1992)\citenamefont {Thole}, \citenamefont {Carra}, \citenamefont {Sette},\ and\ \citenamefont {van~der Laan}}]{Thole1992}%
  \BibitemOpen
  \bibfield  {author} {\bibinfo {author} {\bibfnamefont {B.~T.}\ \bibnamefont {Thole}}, \bibinfo {author} {\bibfnamefont {P.}~\bibnamefont {Carra}}, \bibinfo {author} {\bibfnamefont {F.}~\bibnamefont {Sette}},\ and\ \bibinfo {author} {\bibfnamefont {G.}~\bibnamefont {van~der Laan}},\ }\bibfield  {title} {\bibinfo {title} {{X}-ray circular dichroism as a probe of orbital magnetization},\ }\href {https://doi.org/10.1103/PhysRevLett.68.1943} {\bibfield  {journal} {\bibinfo  {journal} {Phys. Rev. Lett.}\ }\textbf {\bibinfo {volume} {68}},\ \bibinfo {pages} {1943} (\bibinfo {year} {1992})}\BibitemShut {NoStop}%
\bibitem [{\citenamefont {Stout}\ and\ \citenamefont {Reed}(1954)}]{Stout1954}%
  \BibitemOpen
  \bibfield  {author} {\bibinfo {author} {\bibfnamefont {J.~W.}\ \bibnamefont {Stout}}\ and\ \bibinfo {author} {\bibfnamefont {S.~A.}\ \bibnamefont {Reed}},\ }\bibfield  {title} {\bibinfo {title} {The crystal structure of mnf2, fef2, cof2, nif2 and znf2},\ }\href {https://doi.org/10.1021/ja01650a005} {\bibfield  {journal} {\bibinfo  {journal} {Journal of the American Chemical Society}\ }\textbf {\bibinfo {volume} {76}},\ \bibinfo {pages} {5279} (\bibinfo {year} {1954})}\BibitemShut {NoStop}%
\bibitem [{\citenamefont {Blaha}\ \emph {et~al.}()\citenamefont {Blaha}, \citenamefont {Schwarz}, \citenamefont {Madsen}, \citenamefont {Kvasnicka},\ and\ \citenamefont {Luitz}}]{wien2k}%
  \BibitemOpen
  \bibfield  {author} {\bibinfo {author} {\bibfnamefont {P.}~\bibnamefont {Blaha}}, \bibinfo {author} {\bibfnamefont {K.}~\bibnamefont {Schwarz}}, \bibinfo {author} {\bibfnamefont {G.}~\bibnamefont {Madsen}}, \bibinfo {author} {\bibfnamefont {D.}~\bibnamefont {Kvasnicka}},\ and\ \bibinfo {author} {\bibfnamefont {J.}~\bibnamefont {Luitz}},\ }\href@noop {} {\emph {\bibinfo {title} {WIEN2k, An Augmented Plane Wave + Local Orbitals Program for Calculating Crystal Properties (Karlheinz Schwarz, Techn. Universitat Wien, Austria, 2001), ISBN 3-9501031-1-2}}}\BibitemShut {NoStop}%
\bibitem [{\citenamefont {Mostofi}\ \emph {et~al.}(2014)\citenamefont {Mostofi}, \citenamefont {Yates}, \citenamefont {Pizzi}, \citenamefont {Lee}, \citenamefont {Souza}, \citenamefont {Vanderbilt},\ and\ \citenamefont {Marzari}}]{wannier90}%
  \BibitemOpen
  \bibfield  {author} {\bibinfo {author} {\bibfnamefont {A.~A.}\ \bibnamefont {Mostofi}}, \bibinfo {author} {\bibfnamefont {J.~R.}\ \bibnamefont {Yates}}, \bibinfo {author} {\bibfnamefont {G.}~\bibnamefont {Pizzi}}, \bibinfo {author} {\bibfnamefont {Y.-S.}\ \bibnamefont {Lee}}, \bibinfo {author} {\bibfnamefont {I.}~\bibnamefont {Souza}}, \bibinfo {author} {\bibfnamefont {D.}~\bibnamefont {Vanderbilt}},\ and\ \bibinfo {author} {\bibfnamefont {N.}~\bibnamefont {Marzari}},\ }\bibfield  {title} {\bibinfo {title} {An updated version of wannier90: {A} tool for obtaining maximally-localised {Wannier} functions},\ }\href {https://doi.org/http://dx.doi.org/10.1016/j.cpc.2014.05.003} {\bibfield  {journal} {\bibinfo  {journal} {Comput. Phys. Commun.}\ }\textbf {\bibinfo {volume} {185}},\ \bibinfo {pages} {2309 } (\bibinfo {year} {2014})}\BibitemShut {NoStop}%
\bibitem [{\citenamefont {Kune\v{s}}\ \emph {et~al.}(2010)\citenamefont {Kune\v{s}}, \citenamefont {Arita}, \citenamefont {Wissgott}, \citenamefont {Toschi}, \citenamefont {Ikeda},\ and\ \citenamefont {Held}}]{wien2wannier}%
  \BibitemOpen
  \bibfield  {author} {\bibinfo {author} {\bibfnamefont {J.}~\bibnamefont {Kune\v{s}}}, \bibinfo {author} {\bibfnamefont {R.}~\bibnamefont {Arita}}, \bibinfo {author} {\bibfnamefont {P.}~\bibnamefont {Wissgott}}, \bibinfo {author} {\bibfnamefont {A.}~\bibnamefont {Toschi}}, \bibinfo {author} {\bibfnamefont {H.}~\bibnamefont {Ikeda}},\ and\ \bibinfo {author} {\bibfnamefont {K.}~\bibnamefont {Held}},\ }\bibfield  {title} {\bibinfo {title} {Wien2wannier: From linearized augmented plane waves to maximally localized {Wannier} functions},\ }\href {https://doi.org/http://dx.doi.org/10.1016/j.cpc.2010.08.005} {\bibfield  {journal} {\bibinfo  {journal} {Comput. Phys. Commun.}\ }\textbf {\bibinfo {volume} {181}},\ \bibinfo {pages} {1888 } (\bibinfo {year} {2010})}\BibitemShut {NoStop}%
\bibitem [{\citenamefont {de~Groot}\ \emph {et~al.}(1990)\citenamefont {de~Groot}, \citenamefont {Fuggle}, \citenamefont {Thole},\ and\ \citenamefont {Sawatzky}}]{Groot90}%
  \BibitemOpen
  \bibfield  {author} {\bibinfo {author} {\bibfnamefont {F.~M.~F.}\ \bibnamefont {de~Groot}}, \bibinfo {author} {\bibfnamefont {J.~C.}\ \bibnamefont {Fuggle}}, \bibinfo {author} {\bibfnamefont {B.~T.}\ \bibnamefont {Thole}},\ and\ \bibinfo {author} {\bibfnamefont {G.~A.}\ \bibnamefont {Sawatzky}},\ }\bibfield  {title} {\bibinfo {title} {2$p$ {X}-ray absorption of 3$d$ transition-metal compounds: {A}n atomic multiplet description including the crystal field},\ }\href {https://doi.org/10.1103/PhysRevB.42.5459} {\bibfield  {journal} {\bibinfo  {journal} {Phys. Rev. B}\ }\textbf {\bibinfo {volume} {42}},\ \bibinfo {pages} {5459} (\bibinfo {year} {1990})}\BibitemShut {NoStop}%
\end{thebibliography}%


\begin{thebibliography}{12}%
\makeatletter
\providecommand \@ifxundefined [1]{%
 \@ifx{#1\undefined}
}%
\providecommand \@ifnum [1]{%
 \ifnum #1\expandafter \@firstoftwo
 \else \expandafter \@secondoftwo
 \fi
}%
\providecommand \@ifx [1]{%
 \ifx #1\expandafter \@firstoftwo
 \else \expandafter \@secondoftwo
 \fi
}%
\providecommand \natexlab [1]{#1}%
\providecommand \enquote  [1]{``#1''}%
\providecommand \bibnamefont  [1]{#1}%
\providecommand \bibfnamefont [1]{#1}%
\providecommand \citenamefont [1]{#1}%
\providecommand \href@noop [0]{\@secondoftwo}%
\providecommand \href [0]{\begingroup \@sanitize@url \@href}%
\providecommand \@href[1]{\@@startlink{#1}\@@href}%
\providecommand \@@href[1]{\endgroup#1\@@endlink}%
\providecommand \@sanitize@url [0]{\catcode `\\12\catcode `\$12\catcode `\&12\catcode `\#12\catcode `\^12\catcode `\_12\catcode `\%12\relax}%
\providecommand \@@startlink[1]{}%
\providecommand \@@endlink[0]{}%
\providecommand \url  [0]{\begingroup\@sanitize@url \@url }%
\providecommand \@url [1]{\endgroup\@href {#1}{\urlprefix }}%
\providecommand \urlprefix  [0]{URL }%
\providecommand \Eprint [0]{\href }%
\providecommand \doibase [0]{http://dx.doi.org/}%
\providecommand \selectlanguage [0]{\@gobble}%
\providecommand \bibinfo  [0]{\@secondoftwo}%
\providecommand \bibfield  [0]{\@secondoftwo}%
\providecommand \translation [1]{[#1]}%
\providecommand \BibitemOpen [0]{}%
\providecommand \bibitemStop [0]{}%
\providecommand \bibitemNoStop [0]{.\EOS\space}%
\providecommand \EOS [0]{\spacefactor3000\relax}%
\providecommand \BibitemShut  [1]{\csname bibitem#1\endcsname}%
\let\auto@bib@innerbib\@empty
\bibitem [{\citenamefont {Blaha}\ \emph {et~al.}()\citenamefont {Blaha}, \citenamefont {Schwarz}, \citenamefont {Madsen}, \citenamefont {Kvasnicka},\ and\ \citenamefont {Luitz}}]{wien2k}%
  \BibitemOpen
  \bibfield  {author} {\bibinfo {author} {\bibfnamefont {P.}~\bibnamefont {Blaha}}, \bibinfo {author} {\bibfnamefont {K.}~\bibnamefont {Schwarz}}, \bibinfo {author} {\bibfnamefont {G.}~\bibnamefont {Madsen}}, \bibinfo {author} {\bibfnamefont {D.}~\bibnamefont {Kvasnicka}}, \ and\ \bibinfo {author} {\bibfnamefont {J.}~\bibnamefont {Luitz}},\ }\href@noop {} {\emph {\bibinfo {title} {WIEN2k, An Augmented Plane Wave + Local Orbitals Program for Calculating Crystal Properties (Karlheinz Schwarz, Techn. Universitat Wien, Austria, 2001), ISBN 3-9501031-1-2}}}\BibitemShut {NoStop}%
\bibitem [{\citenamefont {Stout}\ and\ \citenamefont {Reed}(1954)}]{Stout1954}%
  \BibitemOpen
  \bibfield  {author} {\bibinfo {author} {\bibfnamefont {J.~W.}\ \bibnamefont {Stout}}\ and\ \bibinfo {author} {\bibfnamefont {S.~A.}\ \bibnamefont {Reed}},\ }\href {\doibase 10.1021/ja01650a005} {\bibfield  {journal} {\bibinfo  {journal} {Journal of the American Chemical Society}\ }\textbf {\bibinfo {volume} {76}},\ \bibinfo {pages} {5279} (\bibinfo {year} {1954})}\BibitemShut {NoStop}%
\bibitem [{\citenamefont {Mostofi}\ \emph {et~al.}(2014)\citenamefont {Mostofi}, \citenamefont {Yates}, \citenamefont {Pizzi}, \citenamefont {Lee}, \citenamefont {Souza}, \citenamefont {Vanderbilt},\ and\ \citenamefont {Marzari}}]{wannier90}%
  \BibitemOpen
  \bibfield  {author} {\bibinfo {author} {\bibfnamefont {A.~A.}\ \bibnamefont {Mostofi}}, \bibinfo {author} {\bibfnamefont {J.~R.}\ \bibnamefont {Yates}}, \bibinfo {author} {\bibfnamefont {G.}~\bibnamefont {Pizzi}}, \bibinfo {author} {\bibfnamefont {Y.-S.}\ \bibnamefont {Lee}}, \bibinfo {author} {\bibfnamefont {I.}~\bibnamefont {Souza}}, \bibinfo {author} {\bibfnamefont {D.}~\bibnamefont {Vanderbilt}}, \ and\ \bibinfo {author} {\bibfnamefont {N.}~\bibnamefont {Marzari}},\ }\href {\doibase http://dx.doi.org/10.1016/j.cpc.2014.05.003} {\bibfield  {journal} {\bibinfo  {journal} {Comput. Phys. Commun.}\ }\textbf {\bibinfo {volume} {185}},\ \bibinfo {pages} {2309 } (\bibinfo {year} {2014})}\BibitemShut {NoStop}%
\bibitem [{\citenamefont {Kune\v{s}}\ \emph {et~al.}(2010)\citenamefont {Kune\v{s}}, \citenamefont {Arita}, \citenamefont {Wissgott}, \citenamefont {Toschi}, \citenamefont {Ikeda},\ and\ \citenamefont {Held}}]{wien2wannier}%
  \BibitemOpen
  \bibfield  {author} {\bibinfo {author} {\bibfnamefont {J.}~\bibnamefont {Kune\v{s}}}, \bibinfo {author} {\bibfnamefont {R.}~\bibnamefont {Arita}}, \bibinfo {author} {\bibfnamefont {P.}~\bibnamefont {Wissgott}}, \bibinfo {author} {\bibfnamefont {A.}~\bibnamefont {Toschi}}, \bibinfo {author} {\bibfnamefont {H.}~\bibnamefont {Ikeda}}, \ and\ \bibinfo {author} {\bibfnamefont {K.}~\bibnamefont {Held}},\ }\href {\doibase http://dx.doi.org/10.1016/j.cpc.2010.08.005} {\bibfield  {journal} {\bibinfo  {journal} {Comput. Phys. Commun.}\ }\textbf {\bibinfo {volume} {181}},\ \bibinfo {pages} {1888 } (\bibinfo {year} {2010})}\BibitemShut {NoStop}%
\bibitem [{\citenamefont {Moriya}(1960)}]{Moriya1960}%
  \BibitemOpen
  \bibfield  {author} {\bibinfo {author} {\bibfnamefont {T.}~\bibnamefont {Moriya}},\ }\href {\doibase 10.1103/PhysRev.117.635} {\bibfield  {journal} {\bibinfo  {journal} {Phys. Rev.}\ }\textbf {\bibinfo {volume} {117}},\ \bibinfo {pages} {635} (\bibinfo {year} {1960})}\BibitemShut {NoStop}%
\bibitem [{\citenamefont {Hariki}\ \emph {et~al.}(2017)\citenamefont {Hariki}, \citenamefont {Uozumi},\ and\ \citenamefont {Kune\ifmmode~\check{s}\else \v{s}\fi{}}}]{Hariki2017}%
  \BibitemOpen
  \bibfield  {author} {\bibinfo {author} {\bibfnamefont {A.}~\bibnamefont {Hariki}}, \bibinfo {author} {\bibfnamefont {T.}~\bibnamefont {Uozumi}}, \ and\ \bibinfo {author} {\bibfnamefont {J.}~\bibnamefont {Kune\ifmmode~\check{s}\else \v{s}\fi{}}},\ }\href {\doibase 10.1103/PhysRevB.96.045111} {\bibfield  {journal} {\bibinfo  {journal} {Phys. Rev. B}\ }\textbf {\bibinfo {volume} {96}},\ \bibinfo {pages} {045111} (\bibinfo {year} {2017})}\BibitemShut {NoStop}%
\bibitem [{\citenamefont {Hariki}\ \emph {et~al.}(2020)\citenamefont {Hariki}, \citenamefont {Winder}, \citenamefont {Uozumi},\ and\ \citenamefont {Kune\ifmmode~\check{s}\else \v{s}\fi{}}}]{Hariki20}%
  \BibitemOpen
  \bibfield  {author} {\bibinfo {author} {\bibfnamefont {A.}~\bibnamefont {Hariki}}, \bibinfo {author} {\bibfnamefont {M.}~\bibnamefont {Winder}}, \bibinfo {author} {\bibfnamefont {T.}~\bibnamefont {Uozumi}}, \ and\ \bibinfo {author} {\bibfnamefont {J.}~\bibnamefont {Kune\ifmmode~\check{s}\else \v{s}\fi{}}},\ }\href {\doibase 10.1103/PhysRevB.101.115130} {\bibfield  {journal} {\bibinfo  {journal} {Phys. Rev. B}\ }\textbf {\bibinfo {volume} {101}},\ \bibinfo {pages} {115130} (\bibinfo {year} {2020})}\BibitemShut {NoStop}%
\bibitem [{\citenamefont {Matarrese}\ and\ \citenamefont {Stout}(1954)}]{Matarrese1954}%
  \BibitemOpen
  \bibfield  {author} {\bibinfo {author} {\bibfnamefont {L.~M.}\ \bibnamefont {Matarrese}}\ and\ \bibinfo {author} {\bibfnamefont {J.~W.}\ \bibnamefont {Stout}},\ }\href {\doibase 10.1103/PhysRev.94.1792} {\bibfield  {journal} {\bibinfo  {journal} {Phys. Rev.}\ }\textbf {\bibinfo {volume} {94}},\ \bibinfo {pages} {1792} (\bibinfo {year} {1954})}\BibitemShut {NoStop}%
\bibitem [{\citenamefont {Hariki}\ \emph {et~al.}(2024)\citenamefont {Hariki}, \citenamefont {Takahashi},\ and\ \citenamefont {Kune\ifmmode~\check{s}\else \v{s}\fi{}}}]{Hariki2024a}%
  \BibitemOpen
  \bibfield  {author} {\bibinfo {author} {\bibfnamefont {A.}~\bibnamefont {Hariki}}, \bibinfo {author} {\bibfnamefont {Y.}~\bibnamefont {Takahashi}}, \ and\ \bibinfo {author} {\bibfnamefont {J.}~\bibnamefont {Kune\ifmmode~\check{s}\else \v{s}\fi{}}},\ }\href {\doibase 10.1103/PhysRevB.109.094413} {\bibfield  {journal} {\bibinfo  {journal} {Phys. Rev. B}\ }\textbf {\bibinfo {volume} {109}},\ \bibinfo {pages} {094413} (\bibinfo {year} {2024})}\BibitemShut {NoStop}%
\bibitem [{\citenamefont {Winder}\ \emph {et~al.}(2020)\citenamefont {Winder}, \citenamefont {Hariki},\ and\ \citenamefont {Kune\ifmmode~\check{s}\else \v{s}\fi{}}}]{Winder20}%
  \BibitemOpen
  \bibfield  {author} {\bibinfo {author} {\bibfnamefont {M.}~\bibnamefont {Winder}}, \bibinfo {author} {\bibfnamefont {A.}~\bibnamefont {Hariki}}, \ and\ \bibinfo {author} {\bibfnamefont {J.}~\bibnamefont {Kune\ifmmode~\check{s}\else \v{s}\fi{}}},\ }\href {\doibase 10.1103/PhysRevB.102.085155} {\bibfield  {journal} {\bibinfo  {journal} {Phys. Rev. B}\ }\textbf {\bibinfo {volume} {102}},\ \bibinfo {pages} {085155} (\bibinfo {year} {2020})}\BibitemShut {NoStop}%
\bibitem [{\citenamefont {Yamaguchi}\ \emph {et~al.}(2024)\citenamefont {Yamaguchi}, \citenamefont {Higashi}, \citenamefont {Regoutz}, \citenamefont {Takahashi}, \citenamefont {Lazemi}, \citenamefont {Che}, \citenamefont {de~Groot},\ and\ \citenamefont {Hariki}}]{Yamaguchi24}%
  \BibitemOpen
  \bibfield  {author} {\bibinfo {author} {\bibfnamefont {T.}~\bibnamefont {Yamaguchi}}, \bibinfo {author} {\bibfnamefont {K.}~\bibnamefont {Higashi}}, \bibinfo {author} {\bibfnamefont {A.}~\bibnamefont {Regoutz}}, \bibinfo {author} {\bibfnamefont {Y.}~\bibnamefont {Takahashi}}, \bibinfo {author} {\bibfnamefont {M.}~\bibnamefont {Lazemi}}, \bibinfo {author} {\bibfnamefont {Q.}~\bibnamefont {Che}}, \bibinfo {author} {\bibfnamefont {F.~M.~F.}\ \bibnamefont {de~Groot}}, \ and\ \bibinfo {author} {\bibfnamefont {A.}~\bibnamefont {Hariki}},\ }\href {\doibase 10.1103/PhysRevB.109.205143} {\bibfield  {journal} {\bibinfo  {journal} {Phys. Rev. B}\ }\textbf {\bibinfo {volume} {109}},\ \bibinfo {pages} {205143} (\bibinfo {year} {2024})}\BibitemShut {NoStop}%
\bibitem [{\citenamefont {Okada}\ and\ \citenamefont {Kotani}(1991)}]{Okada1991}%
  \BibitemOpen
  \bibfield  {author} {\bibinfo {author} {\bibfnamefont {K.}~\bibnamefont {Okada}}\ and\ \bibinfo {author} {\bibfnamefont {A.}~\bibnamefont {Kotani}},\ }\href {\doibase 10.1143/JPSJ.60.772} {\bibfield  {journal} {\bibinfo  {journal} {J. Phys. Soc. Japan}\ }\textbf {\bibinfo {volume} {60}},\ \bibinfo {pages} {772} (\bibinfo {year} {1991})}\BibitemShut {NoStop}%
\end{thebibliography}%
\end{document}


\title{Supplementary Material to ``Separating Altermagnetic and Ferromagnetic Effects in X-ray Magnetic Dichroism of Rutile NiF$_2$''}

\author{A.~Hariki}
\affiliation{Department of Physics and Electronics, Graduate School of Engineering,
Osaka Metropolitan University, 1-1 Gakuen-cho, Nakaku, Sakai, Osaka 599-8531, Japan}
\author{K.~Sakurai}
\affiliation{Department of Physics and Electronics, Graduate School of Engineering,
Osaka Metropolitan University, 1-1 Gakuen-cho, Nakaku, Sakai, Osaka 599-8531, Japan}
\author{T.~Okauchi}
\affiliation{Department of Physics and Electronics, Graduate School of Engineering,
Osaka Metropolitan University, 1-1 Gakuen-cho, Nakaku, Sakai, Osaka 599-8531, Japan}
\author{J.~Kune\v{s}}
\affiliation{Department of Condensed Matter Physics, Faculty of
  Science, Masaryk University, Kotl\'a\v{r}sk\'a 2, 611 37 Brno,
  Czechia}

\maketitle
\begin{figure}[b]
\includegraphics[width=0.4\columnwidth]{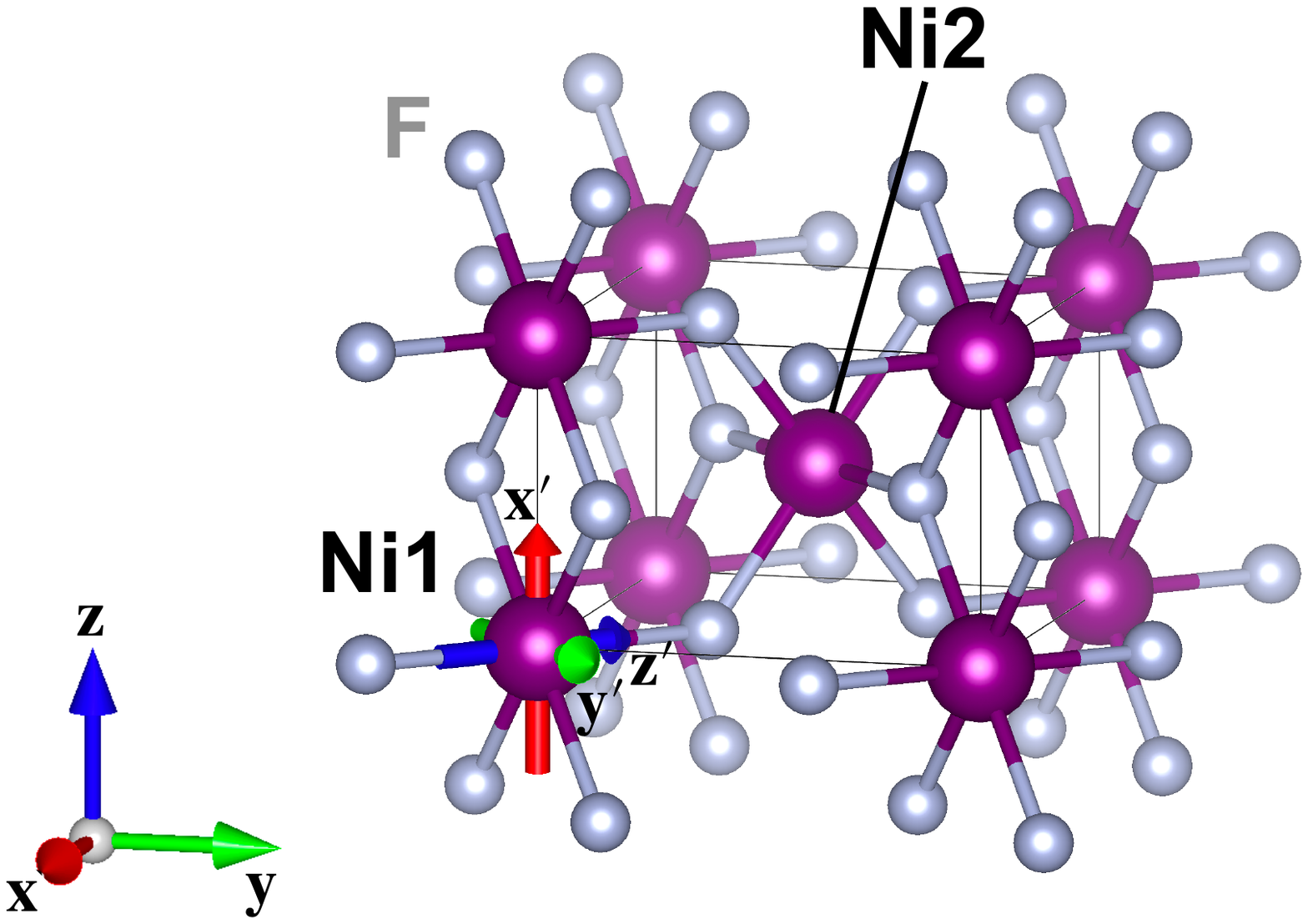}
    \caption{The crystal structure of NiF$_2$ and a local coordinate (x'y'z').}
\label{fig_struct_local}
\end{figure}
\section{Mean-field calculation and XMCD simulation}

Ni $L$-edge x-ray magnetic circular dichroism (XMCD) spectra were calculated based on the Ni$^{2+}$ atomic model with the Weiss field $\bf b$ in the magnetically ordered state, as determined within a mean-field calculation for the spin exchange interaction.  First, density functional theory (DFT) calculations were performed using the WIEN2K package~\cite{wien2k} within the local density approximation for the exchange-correlation potential. The experimental rutile structure~\cite{Stout1954} was employed in the DFT calculations. The crystal-field term, $h^{\text{CF}}$, in the atomic model Hamiltonian was derived from the DFT band structure by projecting the Ni 3$d$ bands near the Fermi level onto a tight-binding model using the wannier90 and wien2wannier packages~\cite{wannier90,wien2wannier}. The $h^{\text{CF}}$ term is expressed in the crystal-field basis ($x'y'$, $3z'^2-r'^2$, $x'^2-y'^2$, $z'x'$, $y'z'$) with respect to the local coordinate system ($x'y'z'$) shown in Fig.~S\ref{fig_struct_local} as,
\begin{equation*}
h^{\text{CF}}=
\begin{pmatrix}
-0.107 & 0 & 0 & 0 & 0 \\
0 & -0.141 & -0.043 & 0 & 0 \\
0 & -0.043 & -1.058 & 0 & 0 \\
0 & 0 & 0 & -0.954 & 0 \\
0 & 0 & 0 & 0 & -0.828 \\
\end{pmatrix}.
\end{equation*}

The $d$--$d$ interaction within the Ni$^{2+}$ configuration can be parameterized by the Hubbard $U = F_0$ and Hund’s coupling $J_{\rm H} = (F^{dd}_2 + F^{dd}_4)/14$. In the atomic model with a fixed number of Ni 3$d$ electrons (3$d^8$), the monopole term $F_0$ contributes only to an overall energy shift and is therefore irrelevant to our analysis. The Hund's coupling parameter is set to $J_{\rm H} = 0.86$ eV, a typical value for Ni$^{2+}$ systems.

\begin{figure}
\includegraphics[width=0.35\columnwidth]{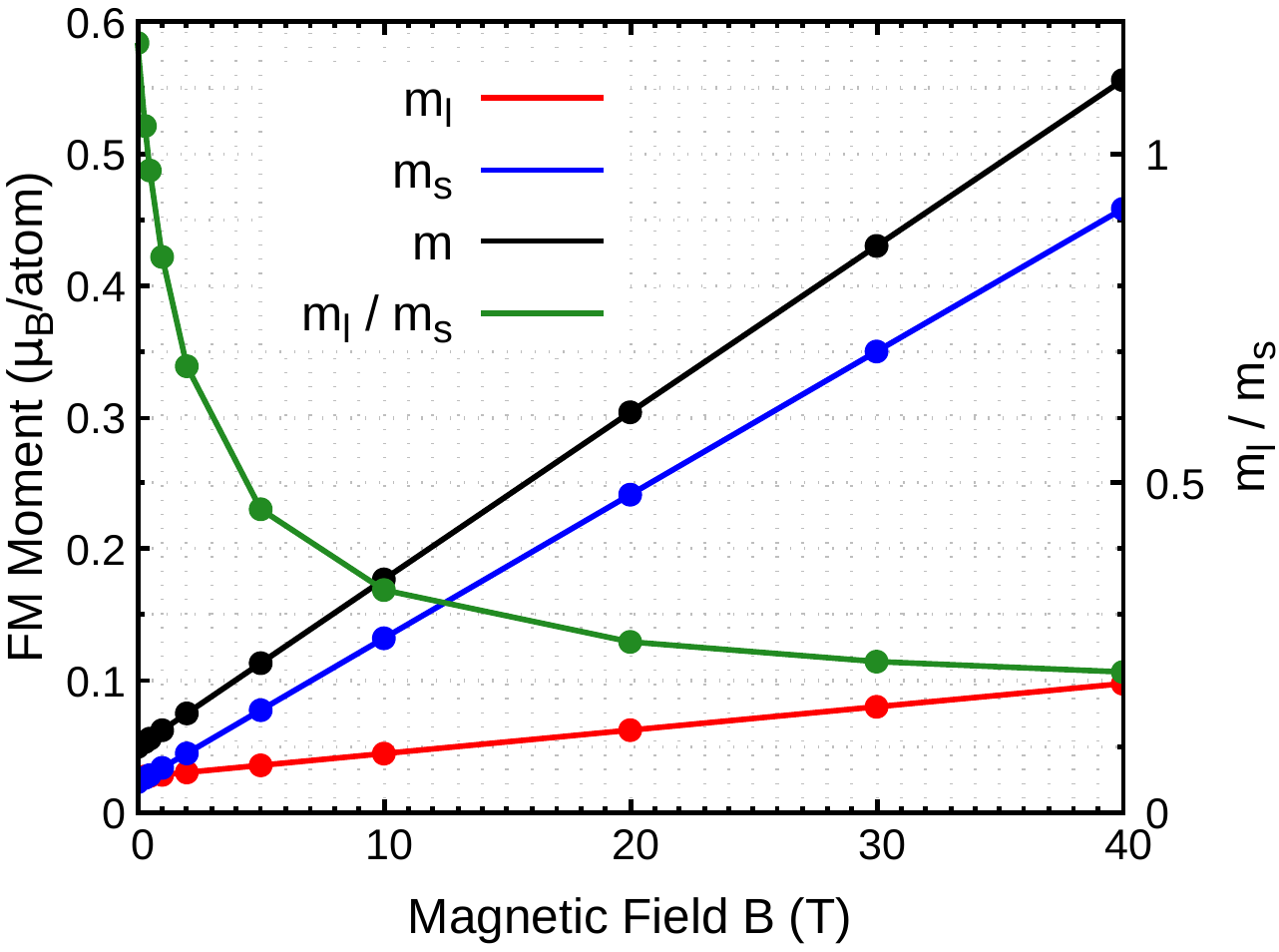}
\caption{Total ($m$), orbital ($m_l$), and spin ($m_s$) components of the ferromagnetic moment, calculated independently for each magnetic field $\bf B$ at $\varphi_B=0^\circ$. The ratio $m_l/m_s$ is shown on the right vertical axis.}
\label{fig_ms_ml_ratio}
\end{figure}

To simulate the magnetically ordered state, we take into account the antiferromagnetic exchange interaction between nearest-neighbor Ni spins of $S=1$ based on the spin Heisenberg Hamiltonian $\mathcal{H}=J\sum_{\langle i,j\rangle}{{\bf S_i\cdot{\bf S}_j}}$ with $J=1.47$~meV~\cite{Moriya1960}. The exchange interaction is treated within a static mean-field approximation. In the self-consistent loop for determining the Weiss field ${\bf b}$, which represents the effective field from surrounding spins, the local spin expectation values $\langle {\bf S}\rangle$ are computed from the Ni$^{2+}$ atomic model described above, explicitly incorporating all 45 multiplet states in the 3$d^8$ configuration. When an external field $\bf B$ is applied, it acts on both the Ni spin (${\bf m}_s$) and the orbital (${\bf m}_l$) magnetic moments at the Ni sites. 
The spin-orbit coupling (SOC) on the Ni 3$d$ shell is included in the mean-field calculation, as well as in the XMCD simulations described below. The SOC constants, $\xi = 77$ meV and 83 meV, were obtained from the DFT calculation and atomic Hartree-Fock calculations~\cite{Hariki2017,Hariki20}, respectively.
Table~I summarizes the Ni ferromagnetic components and the corresponding Weiss field ${\bf b}$ in the absence of the external field $\bf B$ for a wide range of SOC values. 
In the solutions in Table~I, the altermagnetic Néel vector {\bf L} is aligned along the [010] direction ($\varphi_L = 90^\circ$).
To account for hybridization effects with ligands, as explained below, the SOC value was reduced to $\xi = 50$~meV in the presented results.
With $\xi = 50$~meV, the calculated magnetic moment is $m = 0.050~\mu_B$, which agrees reasonably well with previous estimates~\cite{Matarrese1954,Moriya1960}. In Ref.~\cite{Moriya1960}, a value of $\xi = 37$~meV was used. As discussed in Sec.~III, our main conclusion regarding the XMCD properties remains unaffected by the specific choice of the SOC constant. 
Figure~S\ref{fig_ms_ml_ratio} shows the evolution of the magnetic moments computed with $\xi = 50$~meV under the external magnetic field $\bf B$ applied along the [100] axis ($\varphi_B = 0^{\circ}$). As discussed in the main text, the orbital moment contribution is significant in the absence of the external field but is reduced with the field, as reflected in the ratio of $m_l/m_s$ shown in Fig.~S\ref{fig_ms_ml_ratio}.

      \begin{table}[h]
            \centering  
            \renewcommand{\arraystretch}{1.8}
            \setlength{\tabcolsep}{12pt}
            \begin{tabular}{c | c c c c c c c }
            \hline\hline
            $\xi$ (meV) & 100 & 83 & 77 & 50 & 37 & 35 & 30 \\
            \hline
            $m_s$ ($\mu_B$) & 0.087 & 0.061 & 0.053 & 0.023 & 0.013 & 0.012 & 0.009 \\
            $m_l$ ($\mu_B$) & 0.063 & 0.050 & 0.045 & 0.027 & 0.020 & 0.018 & 0.016 \\
            $m$ ($\mu_B$)  & 0.150 & 0.111 & 0.098 & 0.050 & 0.033 & 0.030 & 0.025 \\
            \hline
            $b_{[100]}$ (meV)  & $-$0.257 & $-$0.181 & $-$0.155 & $-$0.067 & $-$0.038 & $-$0.034 & $-$0.025 \\  
            $b_{[010]}$ (meV)  & 5.784  & 5.818  & 5.829  & 5.864  & 5.875  & 5.877  & 5.880 \\  
            \hline\hline
            \end{tabular}
            \vspace*{+0.3cm}
            \caption{Total ($m$), orbital ($m_l$), and spin ($m_s$) contributions to the ferromagnetic moment (per Ni atom) along the [100] direction, calculated for different values of the SOC constant $\xi$. The Weiss field $\bf b$ acting on the Ni1 site along the [100] and [010] directions is also shown.}
            \label{tab_soc_ferromoment}
        \end{table}        

      \begin{table}[h]
            \centering  
            \begin{tabular}{c | c c c c c c }
            \hline\hline\\[-10pt]
            &{\ }
            $F^{dd}_{2}${\ }  \, & {\ }$F^{dd}_{4}${\ } \, & {\ }$F^{pd}_{2}${\ } \, & {\ }$G^{pd}_{1}${\ } \, & {\ }$G^{pd}_{3}${\ } \, & {\ }$\xi_{2p}${\ }
            \\[+4pt]
            \hline\\[-7pt]
            {\quad}Values (eV){\ }{\quad}&{\ }7.409{\ }&{\ }4.631{\ }&{\ }6.563{\ }&{\ }4.919{\ }& {\ }2.797{\ }& {\ }11.507{\ }\\[+7pt]
            \hline\hline
            \end{tabular}
            \vspace*{+0.2cm}
            \caption{Ni$^{2+}$ atomic Slater integral values for the $3d$--$3d$ (valence-valence) and $2p$--$3d$ (core-valence) interactions, as well as the 2$p$ core-orbital SOC $\xi_{2p}$ values used in the Ni$^{2+}$ atomic model. Here, $F$ and $G$ represent the direct and exchange integrals, respectively.}
            \label{tab_slater}
        \end{table}        

After obtaining the mean-field solution for the spin Heisenberg Hamiltonian $\mathcal{H}$, we compute the XMCD using the atomic model with the Weiss field ${\bf b}$ from the converged solution. At this step, the Ni 2$p$ core orbitals and their interaction with the Ni 3$d$ electrons are considered explicitly. The Slater integrals for the 2$p$--3$d$ interaction ($F^{pd}_{2},G^{pd}_{1},G^{pd}_{3}$) and the 2$p$ core-orbital SOC ($\xi_{2p}$) parameters were estimated using atomic Hartree-Fock calculations, as described in Refs.~\cite{Hariki2017,Hariki20}. These parameter values are listed in Table~\ref{tab_slater}. 
The x-ray absorption spectroscopy (XAS) spectral function is expressed using Fermi's golden rule
\begin{equation}
\label{eq:xas}
\begin{split}
F^{(i)}_{\rm XAS}(\omega_{\rm in})&=-\frac{1}{\pi}\operatorname{Im} \langle g | T^{\dag} \frac{1}{\omega_{\rm in}+E_g-H_i} T| g \rangle. \notag \\
\end{split}
\end{equation}
Here, $|g\rangle$ represents the ground state with energy $E_g$, and $\omega_{\rm in}$ denotes the energy of the incident photon. The operator $T$ corresponds to the electric dipole transition operator for circularly-polarized x-rays. The index $i$ ($i = 1, 2$) indicates the x-ray excited Ni site within the unit cell. The atomic Hamiltonian $H_{i}$ for each Ni site is used to calculate its individual contribution to the spectra~\cite{Hariki2024a,Winder20}. The total XMCD intensities are then obtained by summing the contributions from the two sites.


\section{NiF$_6$ cluster model simulation and spin-orbit coupling dependence}

\begin{figure}[htbp]
\includegraphics[width=0.7\columnwidth]{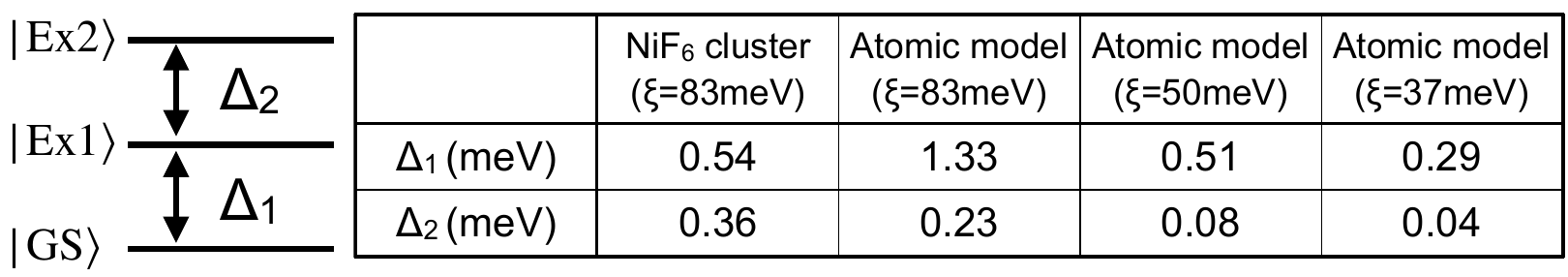}
    \caption{The energy splittings between the ground state and the first excited state ($\Delta_{1}$), as well as between the first and second excited states ($\Delta_{2}$) within the lowest $S=1$ states, are shown for the NiF$_6$ cluster model and the atomic model in meV unit. For the atomic model, simulations were conducted with three different SOC constants $\xi$ (83~meV, 50~meV, 37~meV).}
    \vspace{-0.5cm}
\label{fig_energy}
\end{figure}

In Fig.~S\ref{fig_energy}, we compare the energy splittings of the lowest $S=1$ states in the local excitation spectrum without both the external and exchange fields between the NiF$_6$ cluster model and the Ni$^{2+}$ atomic model. These splittings reflect the single-ion anisotropy. The NiF$_6$ cluster model explicitly includes hybridization with ligands, whereas in the atomic model, this effect is incorporated into the $d$ level energies effectively. The hybridization parameters in the NiF$_6$ cluster model, summarized in Table~II, were derived from a tight-binding model Hamiltonian that explicitly includes both the Ni 3$d$ and F 2$p$ DFT bands (see Ref.~\cite{Yamaguchi24} for details about our construction of the cluster model). The averaged Coulomb interaction $U_{dd}$ and charge-transfer energy $\Delta_{\rm CT}$ in the cluster-model Hamiltonian were adopted from Ref.~\cite{Okada1991}. The cluster model implements $\xi = 83$~meV, obtained from atomic Hartree-Fock calculations. The Ni$^{2+}$ atomic model with $\xi = 50$~meV reproduces a similar splitting, $\Delta_1$, between the lowest and first excited states as obtained in the NiF$_6$ cluster model calculations. Thus, we used the reduced value $\xi = 50$~meV in our atomic model calculations in the main text to account for ligand hybridization effects. 
%
%
      \begin{table}[h]
            \centering  
            \begin{tabular}{c | c c c c c c c c}
            \hline\hline\\[-10pt]
            &{\ }
            \,\,$U_{dd}${\ }  \, & {\ }\,\,$\Delta_{\rm CT}${\ } \, & {\ }\,\,$V_{B_{1g}}${\ } \, & {\ }\,\,$V_{A_{g}}${\ } \, & {\ }\,\,$V_{A'_{g}}${\ } \, & {\ }\,\,$V_{B_{2g}}${\ } \, & {\ }\,\,$V_{B_{3g}}${\ }
            \\[+4pt]
            \hline\\[-7pt]
            {\quad}Values (eV){\ }{\quad}&{\ }6.50{\ }&{\ }4.30{\ }&{\ }2.18{\ }&{\ }2.06& {\ }1.19{\ }&{\ }1.18 {\ }&{\ }1.06{\ }\\[+7pt]
            \hline\hline
            \end{tabular}
            \vspace*{+0.2cm}
            \caption{The parameter values for configuration-averaged $3d-3d$ Coulomb interaction ($U_{dd}$), charge-transfer energy ($\Delta_{\rm CT}$) and the metal-ligand hybridization amplitude adopted in the NiF$_6$ cluster model in eV unit. $U_{dd}$ and $\Delta_{\rm CT}$ are taken from Ref.~\cite{Okada1991}.}
            \label{tab_cluster_param}
        \end{table}        
\section{Dependence of the Magnetic State and XMCD on SOC}
\begin{figure}[H]
\begin{center}
\includegraphics[width=0.70\columnwidth]{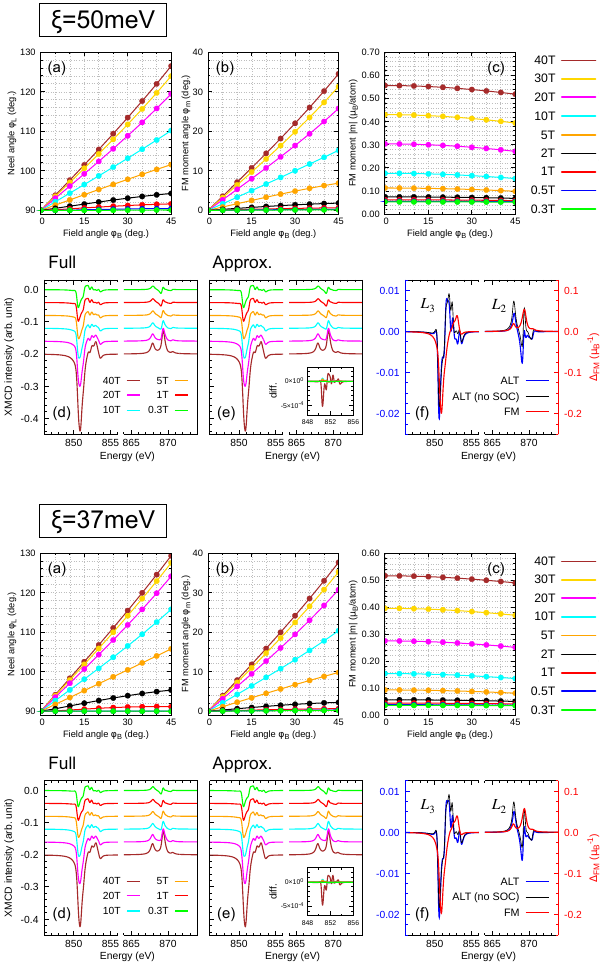}
\end{center}
\caption{Calculated relation (with $\xi=50$~meV and 37~meV) between the angles of the external field $\varphi_B$ and (a) the N\'eel vector $\varphi_L$, (b) the FM moment $\varphi_m$, and (c) the amplitude of the FM moment $|{\bf m(B)}|$ for selected amplitudes of the external field $\bf B$. (d) Ni $L_{2,3}$-edge XMCD intensities in NiF$_2$ calculated independently 
for each magnetic field $B$ and $\varphi_B = 0^\circ$ with no approximations to the method ('FULL').
(e) XMCD intensities computed as a linear combination ('APPROX.') of (f) $\Delta_{\rm ALT}(\omega)$ (blue, left axis) and $\Delta_{\rm FM}(\omega)$ (red, right axis).
$\Delta_{\rm ALT}(\omega)$ in the nonrelativistic limit, calculated without the Ni 3$d$ valence SOC, is also shown (thin black, left axis).
The inset in panel (e) shows the difference in the XMCD intensities at the Ni $L_3$-edge between the full calculations in (d) and the approximations in (e).}
\label{fig:ang_dep_moriya}
\end{figure}

\begin{figure}[H]
\includegraphics[width=1.0\columnwidth]{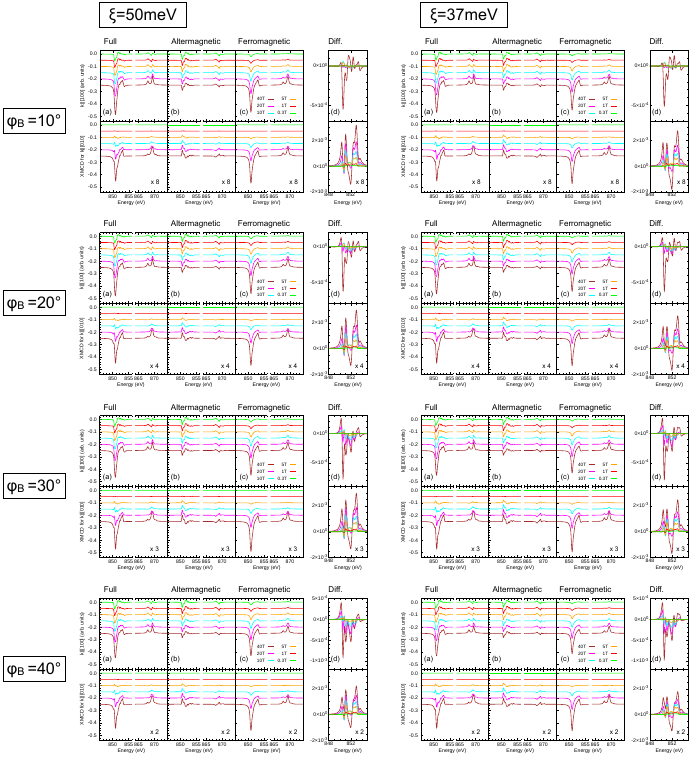}
    \caption{(a) Ni $L_{2,3}$-edge XMCD intensities in NiF$_2$ calculated for various magnetic field amplitudes $B$ with $\varphi_B = 10^\circ$, $20^\circ$, $30^\circ$, and $40^\circ$ (from top to bottom) for the two geometries of the light propagation vector $\hat{\bf k}=[100]$ (top) and $\hat{\bf k}=[010]$ (bottom). The left and right panels show results obtained with $\xi = 50$~meV and $\xi = 37$meV, respectively. (b) Altermagnetic contributions. (c) Ferromagnetic contributions. (d) The difference in XMCD intensities at the Ni $L_3$-edge between the full calculations in (a) and the approximations using Eq.~(1) in the main text.}
\label{fig:xmcd_tilt_sm}
\end{figure}

Figure~S\ref{fig:ang_dep_moriya} summarizes the magnetic state and XMCD properties for different Ni 3$d$ SOC values, $\xi = 50$~meV and $\xi = 37$~meV, respectively. The magnetic moments in the absence of an external field are listed in Table~I. The smaller SOC value, $\xi = 37$~meV, appears to yield magnetic moment values close to existing experimental estimates~\cite{Matarrese1954,Moriya1960}. As expected for weaker SOC, the ferromagnetic moment is reduced, while the tilt of the N\'eel vector $\bf L$ induced by an external field with $\varphi_B \neq 0$ is enhanced slightly compared to the case with $\xi = 50$~meV.
We found that the Ni $L$-edge XMCD properties are very similar for the two SOC values. In particular, the approximation for the XMCD intensities based on Eq.~(1) provides comparable accuracy in both cases, as shown in Fig.~S\ref{fig:ang_dep_moriya}(e). We provide the XMCD data for various $\varphi_B$ values in Fig.~S\ref{fig:xmcd_tilt_sm}.

\section{Analysis of the XMCD sum rule}
\begin{figure}[htbp]
\includegraphics[width=0.75\columnwidth]{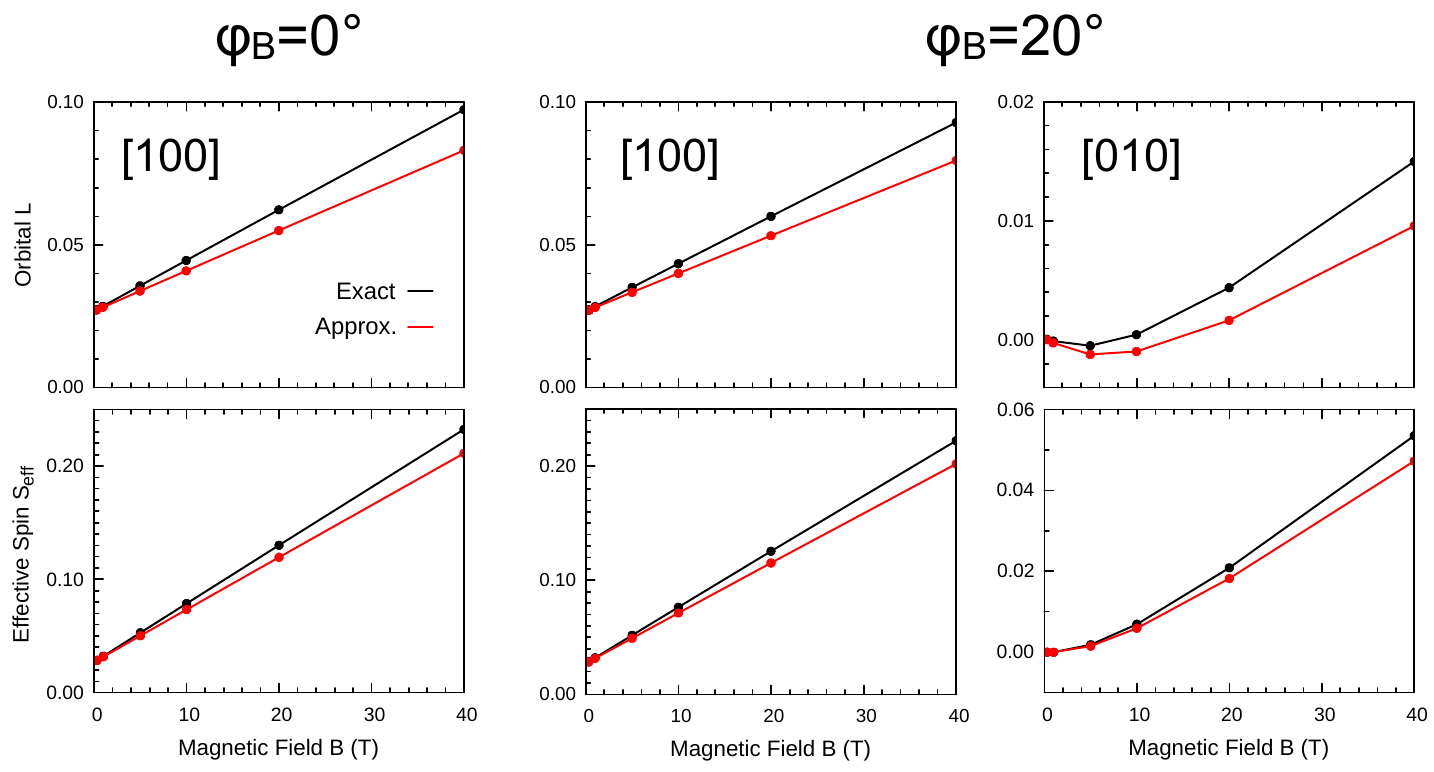}
    \caption{The orbital angular momentum $L$ (top) and effective spin $S_{\rm eff}=S+\frac{7}{2}T$ (bottom) obtained from the exact XMCD simulation (black) and the approximate scheme in Eq.~(1) (red) are shown for $\varphi_B=0^\circ$ (left) and $\varphi_B=20^\circ$ (middle, right) and different amplitudes of the external field $\mathbf{B}$. The middle and right panels correspond to the projections of these quantities onto the [100] and [010] directions, respectively.}
\label{fig:fig_sum_sm}
\end{figure}

Figure~\ref{fig:fig_sum_sm} summarizes the orbital moment ${\bf L}$ and effective spin ${\bf S}_{\rm eff} = {\bf S} + \frac{7}{2} {\bf T}$ derived from the sum rule applied to the simulated XMCD results. The exact values correspond to those obtained from the sum rule for the ``Full" spectra in the main text. We checked numerically that these values are consistent with the expectation values directly evaluated from the atomic-model ground state. The values derived from the approximated XMCD based on Eq.~(1) are compared with the exact values. 
The approximated values accurately reproduce the exact ones in the weak external field $\bf B$, while the error increases somewhat as the field strength increases. For 40~T, deviation of 14.6~\% and 9.11~\% at $\varphi_B=0^{\circ}$ is expected for the orbital and effective spin values, respectively. When the field is rotated to $\varphi_B = 20^\circ$, the deviation for the [100] component remains nearly unchanged (14.4~\% for $\bf L$, 9.16~\% for ${\bf S}_{\rm eff}$). However, for the [010] component, particularly for $\bf L$, it increases (36.1~\% for $\bf L$, 11.7~\% for ${\bf S}_{\rm eff}$).


\bibliography{main}